\documentclass[a4paper,11pt]{article}
\usepackage{jcappub} 
\usepackage{orcidlink}
\usepackage{lineno}

\newcommand{\citetjcap}[2]{#1~\cite{#2}}

\def\be{\begin{equation}}
\def\ee{\end{equation}}
\def\ba{\begin{eqnarray}}
\def\ea{\end{eqnarray}}

\title{A Gaussian Covariance Matrix for Joint Pre- and Post-Reconstruction Full-Shape Power Spectrum Analysis}
\author[1]{{Yuting Wang}\orcidlink{0000-0001-7756-8479},}
\author[2]{{Ruiyang Zhao}\orcidlink{0000-0002-7284-7265},}
\author[1,3]{{Gong-Bo Zhao}\orcidlink{0000-0003-4726-6714},}
\author[4,5,6]{{Kazuya Koyama}\orcidlink{0000-0001-6727-6915},}

\affiliation[1]{National Astronomical Observatories, Chinese Academy of Sciences, Beijing, 100101, P.R.China}
\affiliation[2]{Department of Astronomy, Tsinghua University, Beijing 100084, China}
\affiliation[3]{University of Chinese Academy of Sciences, Beijing, 100049, P.R.China}
\affiliation[4]{Institute of Cosmology and Gravitation, University of Portsmouth, Dennis Sciama Building, Portsmouth, PO1 3FX, UK}
\affiliation[5]{Kavli IPMU (WPI), UTIAS, The University of Tokyo, Kashiwa, Chiba 277-8583, Japan}
\affiliation[6]{Yukawa Institute for Theoretical Physics, Kyoto University, Kyoto 606-8502, Japan}

\emailAdd{ytwang@nao.cas.cn}
\emailAdd{zhaoruiyang@mail.tsinghua.edu.cn}
\emailAdd{gbzhao@nao.cas.cn}
\emailAdd{kazuya.koyama@port.ac.uk}

\abstract{
We apply the Gaussian covariance formalism to develop a semi-analytical covariance model for the joint analysis of pre-reconstruction, post-reconstruction, and cross full-shape galaxy power spectra. We model the reconstruction-reduced, scale-dependent cross shot noise using displacement-field statistics and introduce a new estimator that directly measures this term. Using the measured power spectra and the modeled shot-noise predictions as inputs, we construct the Gaussian covariance while accounting for correlations between the pre- and post-reconstruction density fields. We validate the resulting semi-analytical Gaussian covariance against mock catalogues. Using emulator-based parameter inference, we demonstrate that the semi-analytical Gaussian covariance adequately captures the dominant contribution to the covariance structure of the full data vector ($P_{\ell}^{\rm pre}, P_{\ell}^{\rm post}, P_{\ell}^{\rm cross}$). For the joint fit to these three power spectra, it yields cosmological constraints consistent with those obtained using the mock-based numerical covariance over the adopted fitting ranges: $k_{\rm max}=0.18\,h\,{\rm Mpc}^{-1}$ for $P_{\rm pre}$ and $P_{\rm post}$, and $k_{\rm max}=0.12\,h\,{\rm Mpc}^{-1}$ for $P_{\rm cross}$.
}

\begin{document}


\maketitle

\section{Introduction}

Galaxy redshift surveys provide a powerful probe of cosmology through measurements of the large-scale structure of the Universe. Among two-point clustering statistics, full-shape analyses of the galaxy power spectrum simultaneously exploit information from baryon acoustic oscillations (BAO), redshift-space distortions (RSD), and the broadband shape of the matter power spectrum.

Density-field reconstruction is widely used in galaxy clustering analyses to enhance the BAO signal by partially reversing nonlinear structure evolution \citep{Eisenstein:2006nk}. Traditionally, cosmological analyses combine full-shape information from the pre-reconstruction power spectrum with BAO measurements from the post-reconstruction field \citep{Gil-Marin:2022hnv}. Because both measurements are derived from the same galaxy sample, they are statistically correlated, and their covariance can be estimated either from large suites of mock catalogues or through analytical approximations \citep{Gil-Marin:2022hnv,Maus:2026wsb}.

Beyond this conventional approach, \citetjcap{Wang et al.}{Wang:2022nlx} proposed a joint full-shape analysis of the pre-reconstruction, post-reconstruction, and cross galaxy power spectra, demonstrating that these observables contain complementary cosmological information. This approach has recently been applied to galaxy survey data \citep{DESI:2026ylh}. A key ingredient in such analyses is an accurate covariance matrix that consistently accounts for correlations among all three power-spectrum measurements. Although this covariance can be estimated numerically from large suites of mock catalogues, doing so is computationally expensive, particularly for high-dimensional data vectors. An analytical treatment therefore provides a more efficient alternative \citep{Hikage:2020fte,Zhao:2024xit}.

In this work, we present a semi-analytical Gaussian covariance framework for the joint analysis of the pre-reconstruction, post-reconstruction, and cross full-shape power spectra. The framework combines analytical Gaussian covariance calculations with measured or simulated power spectra as inputs, providing an efficient alternative to mock-based covariance estimation while capturing the dominant covariance structure among the three power-spectrum measurements.

The paper is organized as follows. Sec.~\ref{sec:covmat} presents the Gaussian covariance formalism. Secs.~\ref{sec:xshotnoise} and \ref{sec:xSNestimator} describe the modelling and estimation of the reconstruction-induced cross shot noise, respectively. In Sec.~\ref{sec:result}, we validate the semi-analytical Gaussian covariance against numerical estimates derived from mock catalogues. Finally, we summarize our conclusions in Sec.~\ref{sec:conclusion}.

\section{Gaussian covariance matrix}\label{sec:covmat}

In this section, we adopt the Gaussian covariance formalism for the joint pre-reconstruction, post-reconstruction, and cross power spectra. 

The data vector considered in this work consists of the multipoles of the pre-reconstruction, post-reconstruction, and cross power spectra,
\ba
\mathbf{P} = \left[P_{\ell}^{\rm pre}(k), P_{\ell}^{\rm post}(k), P_{\ell}^{\rm cross}(k)\right]^{T}\,.
\label{eq:Pvector}
\ea
Since these three power spectra are measured from correlated density fields, the corresponding covariance matrix must include all auto- and cross-correlations among the three observables. Namely, the covariance matrix can be written as 
\begin{equation}
{\rm C}_{\ell\ell'} (k)=
\begin{pmatrix}
{\rm C}^{\rm pre,pre}_{\ell\ell'} & {\rm C}^{\rm pre,post}_{\ell\ell'} & {\rm C}^{\rm pre,cross}_{\ell\ell'} \\
{\rm C}^{\rm post,pre}_{\ell\ell'} & {\rm C}^{\rm post,post}_{\ell\ell'} & {\rm C}^{\rm post,cross}_{\ell\ell'} \\
{\rm C}^{\rm cross,pre}_{\ell\ell'} & {\rm C}^{\rm cross,post}_{\ell\ell'} & {\rm C}^{\rm cross,cross}_{\ell\ell'}
\end{pmatrix}\,,
\end{equation}

Assuming Gaussian density fluctuations, the covariance can be computed analytically using Wick's theorem. We first consider two generic power spectra, $P_{ab}$ and $P_{cd}$, constructed from the density fields $a,b,c,$ and $d$. The Gaussian covariance of auto- and cross-power spectra has been extensively studied \cite{Smith:2008ut, Barreira:2020ekm, Montano:2026qmu}. The Gaussian covariance diagonal kernel entering the covariance between $P_{ab}$ and $P_{cd}$ is given by
\begin{equation}
\mathcal{C}_{ab,cd}(k,\mu) = \frac{1}{2} \left[ \widetilde P_{ac}(k,\mu)\widetilde P_{bd}(k,\mu) + \widetilde P_{ad}(k,\mu)\widetilde P_{bc}(k,\mu) \right],
\label{eq:gaussian_kernel}
\end{equation}
where $\mu\equiv\hat{\boldsymbol{k}}\cdot\hat{\boldsymbol{z}}$ is the cosine of the angle between the Fourier wavevector $\boldsymbol{k}$ and the global line-of-sight direction $\hat{\boldsymbol{z}}$. The factor of $1/2$ is included so that the multipole covariance below can be written with the common prefactor $2/N_k$.

For the three types of power spectra in Eq.~(\ref{eq:Pvector}), the Gaussian covariance diagonal-kernel matrix can be written as \citep{White:2008jy, Blake:2013nif}
\begin{equation}
\mathcal{C}(k,\mu)=
\begin{pmatrix}
\widetilde P_{\rm pre}^2 &\widetilde P_{\rm cross}^2 &\widetilde P_{\rm pre}\widetilde P_{\rm cross}\\
\widetilde P_{\rm cross}^2 &\widetilde P_{\rm post}^2 &\widetilde P_{\rm post}\widetilde P_{\rm cross}\\
\widetilde P_{\rm pre}\widetilde P_{\rm cross} &\widetilde P_{\rm post}\widetilde P_{\rm cross} &\frac{1}{2}\left(\widetilde P_{\rm cross}^2+ \widetilde P_{\rm pre}\widetilde P_{\rm post} \right)
\end{pmatrix}. \label{eq:covaniso}
\end{equation}

Here, the 2D power spectrum can be reconstructed from its multipoles as
\begin{equation}
\widetilde P(k,\mu) = \sum_{\ell} \left[ P_{\ell}(k)+N_{\ell}(k) \right]{\cal L}_{\ell}(\mu)\,,
\label{eq:pk3D}
\end{equation}
where $P_{\ell}$ and $N_{\ell} $ denote the shot-noise-subtracted power spectrum multipoles and shot-noise multipoles, and ${\cal L}_{\ell}$ is the Legendre polynomial. 

The covariance of the multipoles of the pre-reconstructed, post-reconstructed, and their cross power spectra is obtained through Legendre projection \citep{Taruya:2010mx}, 
\begin{equation}
\begin{aligned}
{\rm C}_{\ell\ell'} (k) = \frac{2}{N_k}\frac{(2\ell+1)(2\ell'+1)}{2} \int_{-1}^{1} d\mu\, {\cal L}_{\ell}(\mu) {\cal L}_{\ell'}(\mu) \mathcal{C}(k,\mu) 
\,,
\end{aligned}
\end{equation}
where $N_k$ is the number of Fourier modes in the $k$ bin, given by $N_k=(4\pi k^2\Delta k V)/ (2\pi)^3$ with $V$ being the survey volume. 

\section{Cross shot noise modelling}\label{sec:xshotnoise}

The shot-noise contribution in Eq.~(\ref{eq:pk3D}) is different for auto-correlation and cross-power spectra. For the auto-correlation power spectra, $P_{\rm pre}$ and $P_{\rm post}$, we subtract the shot noise by,
\begin{equation}
N^{\rm auto}_{\ell}= (1+\alpha)\frac{1}{\bar n} \delta_{\ell,0}\,,
\end{equation}
which contributes only to the monopole. Here $\alpha$ is the ratio of total number of galaxies to randoms and accounts for the Poisson noise contribution associated with the finite random catalogue. For the periodic-box $P_{\rm pre}$ measurement without a random catalogue, we set $\alpha=0$. For $P_{\rm post}$, we adopt the {\bf RecSym}  convention (where both galaxies and randoms are shifted by the same displacement field) \citep{Chen:2024eri} and use a random catalogue containing 20 times as many objects as the mock galaxy catalogue, corresponding to $\alpha$=0.05.

The shot noise in the $P_{\rm cross}$ is modified by the reconstruction displacement field \citep{Sugiyama:2024eye}. The reconstruction displacement introduces a scale-dependent damping of the shot noise contribution. In this work, we adopt the standard reconstruction algorithm \citep{Eisenstein:2006nk} and retain RSD following the {\bf RecSym} convention \citep{Chen:2024eri}.  In the plane-parallel approximation, the redshift-space displacement in configuration space $\boldsymbol{s}^{z}(\boldsymbol{x})$ is 
\begin{equation}
\boldsymbol{s}^{z}(\boldsymbol{x}) = \boldsymbol{s}^{r}(\boldsymbol{x}) + f_{\rm fid} \left[\boldsymbol{s}^{r}(\boldsymbol{x})\cdot\hat{\boldsymbol z} \right] \hat{\boldsymbol z}\,,
\label{eq:szconfig}
\end{equation}
where $\boldsymbol{s}^{r}(\boldsymbol{x})$ denotes the estimated displacement field in real space, $f_{\rm fid}$ is the fiducial linear growth rate adopted for reconstruction, and we take $\hat{\boldsymbol z}$ to be the global line-of-sight direction.

To compute the statistical moments of the displacement field, it is convenient to work in Fourier space. The Fourier transform of the real-space displacement field is
\begin{equation}
\boldsymbol{s}^{r}(\boldsymbol{q}) = -i \frac{\boldsymbol q}{q^{2}} \frac{K_{s}(q)\, \delta_{g}^{z}(\boldsymbol q)} {b_{\rm fid} + f_{\rm fid}\mu_q^{2}},
\label{eq:srfourier}
\end{equation}
where $\mu_q\equiv \hat{\boldsymbol q}\cdot\hat{\boldsymbol z}$, with $\boldsymbol q$ denoting the integration wavevector used in calculating the displacement-field statistics below. $b_{\rm fid}$ and $f_{\rm fid}$ are the fiducial linear bias and growth rate used in the reconstruction procedure, respectively, and the Gaussian smoothing kernel is $K_s(q) = \exp\left( -q^{2}\Sigma_s^{2}/2 \right)$ with the smoothing scale $\Sigma_s$. 

Combining Eq.~(\ref{eq:szconfig}) with the above equation, the redshift-space displacement field in Fourier space can be written as
\begin{equation}
s_i^z(\boldsymbol{q}) =-i\, \frac{K_s(q)\,\delta_g^z(\boldsymbol{q})} {q^2\left(b_{\rm fid}+f_{\rm fid}\mu_q^2\right)} \left(q_i+f_{\rm fid}q\mu_q\hat{z}_i\right)\,.
\label{eq:szfourier}
\end{equation}
For simplicity, we omit the redshift-space superscript $z$ in the following equations.

The number densities of the discrete density field before and after the reconstruction are
\begin{equation}
n(\boldsymbol{k}) = \sum_{i=1}^{N_{\rm g}} e^{-i\boldsymbol{k}\cdot\boldsymbol{x}_{i}}, ~~~n_{\rm rec}(\boldsymbol{k}) = \sum_{i=1}^{N_{\rm g}} e^{-i\boldsymbol{k}\cdot\boldsymbol{x}_{{\rm rec},i}} = \sum_{i=1}^{N_{\rm g}} e^{-i\boldsymbol{k}\cdot [\boldsymbol{x}_{i}+\boldsymbol{s}(\boldsymbol{x}_{i})]} \,.
\label{eq:numdensity}
\end{equation}
then the cross shot-noise contribution induced by the reconstruction displacement field can be written as
\begin{equation}
N_{\rm cross}(k,\mu) = \left\langle \frac{V}{N_{\rm g}^{2}} \sum_{i=j}^{N_{\rm g}} e^{-i\boldsymbol{k}\cdot \left(\boldsymbol{x}_{i}-\boldsymbol{x}_{{\rm rec}, j}\right)} \right\rangle = \left\langle \frac{V}{N_g^2} \sum_{i=1}^{N_g} e^{i\boldsymbol{k}\cdot\boldsymbol{s}(\boldsymbol{x}_i)} \right\rangle = \frac{1}{\bar n}  \left\langle e^{i\boldsymbol{k}\cdot\boldsymbol{s}(\boldsymbol{x})} \right\rangle \,.
\end{equation}
Using the cumulant expansion theorem, $ \left\langle e^{-i\boldsymbol{k}\cdot\boldsymbol{s}} \right\rangle = \exp\left[ \sum_{n} \frac{(-i)^n}{n!} \left\langle (\boldsymbol{k}\cdot\boldsymbol{s})^n \right\rangle_{\rm c} \right]$,where $\langle\cdots\rangle_{\rm c}$ denotes the connected moments (cumulants). For a statistically homogeneous displacement field with zero mean, $\left\langle \boldsymbol{k}\cdot\boldsymbol{s} \right\rangle_{\rm c}=0$. Assuming that the displacement field is Gaussian, all connected moments higher than second order vanish, $ \left\langle (\boldsymbol{k}\cdot\boldsymbol{s})^n \right\rangle_{\rm c}=0$, for $n>2$.

Therefore, the cumulant expansion is truncated at second order, i.e.\,,
\begin{equation}
\begin{aligned} 
\left\langle e^{-i\boldsymbol{k}\cdot\boldsymbol{s}} \right\rangle = \exp \left[ -\frac{1}{2} \left\langle (\boldsymbol{k}\cdot\boldsymbol{s})^2 \right\rangle_{\rm c} \right]  
= \exp \left[ -\frac{1}{2} k_i k_j \left\langle s_i s_j \right\rangle \right]
= \exp \left[ -\frac12 k_i k_j \kappa_{ij} \right]\,,
\end{aligned}
\end{equation}
where the displacement covariance tensor is defined as $ \kappa_{ij} \equiv \left\langle s_i(\boldsymbol{x})s_j(\boldsymbol{x}) \right\rangle$. Substituting Eq.~(\ref{eq:szfourier}) into this expression and using the definition of the power spectrum, $\left\langle \delta(\boldsymbol q) \delta(\boldsymbol q') \right\rangle = (2\pi)^3 \delta_{\rm D} (\boldsymbol q+\boldsymbol q') P_{gg}(q,\mu_q)$, we obtain
\begin{equation}
\begin{aligned}
\kappa_{ij} = \int \frac{d^{3}q}{(2\pi)^3} \frac {K_s^{2}(q)} {\left( b_{\rm fid} + f_{\rm fid}\mu_q^{2} \right)^2} \frac {P_{gg}(q,\mu_q)} {q^{4}} \times \left[ q_iq_j + f_{\rm fid} q\mu_q (q_i\hat z_j+\hat z_iq_j) + f_{\rm fid}^{2} q^{2}\mu_q^{2} \hat z_i\hat z_j \right]\,.
\end{aligned}
\label{eq:kappa}
\end{equation}

After performing the angular integration of the displacement covariance, the cross shot noise becomes
\begin{equation}
\small
N_{\rm cross}(k,\mu) = \frac{1}{\bar n} \exp \left\{ -\frac12 k^2 \left[ \sigma_A^2(\mu) + (2f_{\rm fid}+f_{\rm fid}^2) \mu^2\sigma_B^2 \right] \right\}\,,
\label{eq:Ncross2D}
\end{equation}
where, 
\begin{equation}
\begin{gathered}
\sigma_A^2(\mu) = \frac{1}{2(2\pi)^2} \int_0^\infty dq \int_{-1}^{1} d\mu_q\, \frac{K_s^2(q)} {\left(b_{\rm fid}+f_{\rm fid}\mu_q^2\right)^2} \times \left[ 1-\mu_q^2 +\left(3\mu_q^2-1\right)\mu^2 \right] P_{gg}(q,\mu_q), \\
\sigma_B^2 = \frac{1}{(2\pi)^2} \int_0^\infty dq \int_{-1}^{1} d\mu_q\, \frac{K_s^2(q)} {\left(b_{\rm fid}+f_{\rm fid}\mu_q^2\right)^2} \times \mu_q^2\, P_{gg}(q,\mu_q)\,,
\end{gathered}
\end{equation}
where $P_{gg}(q,\mu_q)$ is approximated by the Kaiser redshift-space power spectrum, including the Poisson shot-noise contribution,
\ba
P_{gg}(q,\mu_q) = \left(b +f \mu_q^2\right)^2 P_{\rm lin}(z) + \frac{1}{\bar n}\,.
\ea
The corresponding multipole shot noise contribution is obtained through Legendre projection,
\begin{equation}
N^{\rm cross}_\ell(k) = \frac{2\ell+1}{2} \int_{-1}^{1} d\mu\, N_{\rm cross}(k,\mu) {\cal L}_\ell(\mu)\,.
\label{eq:Ncrossell}
\end{equation}

When the fiducial parameters perfectly match the true values, $b_{\rm fid}=b$ and $f_{\rm fid}=f$, and the Poisson contribution
to $P_{gg}$ is neglected, the Kaiser factors cancel, yielding
\begin{equation}
\sigma_A^2=\sigma_B^2\equiv\sigma^2 = \frac{1}{6\pi^2} \int_0^\infty dq\,K_s^2(q)P_{\rm lin}(q).
\end{equation}
Consequently, the cross shot noise reduces to
\begin{equation} 
N_{\rm cross}(k,\mu) = \frac{1}{\bar n} \exp\left[ -\frac{1}{2}k^2\sigma^2 \left\{1+\left(2f+f^2\right)\mu^2\right\} \right].
\end{equation}

\begin{figure*}[htp]
    \centering
    \includegraphics[width=0.9\textwidth]{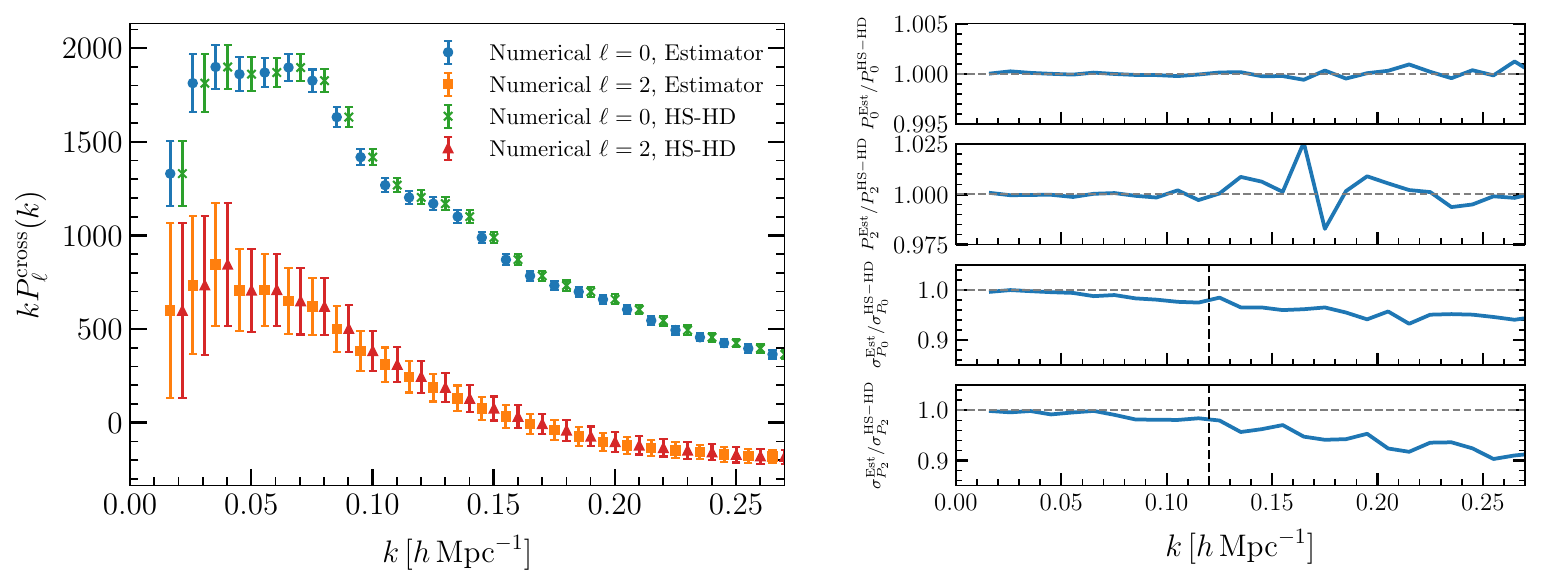}
    \caption{Left: Comparison of the shot-noise-subtracted cross-power spectrum using the new estimator in Eq.~(\ref{eq:xpk2}) and the HS-HD method in Eq.~(\ref{eq:xpk1}). For clarity, the HS-HD-based measurements are horizontally shifted by $0.005\,h\,{\rm Mpc}^{-1}$. Right: Their power-spectrum ratios and variance ratios for the monopole and quadrupole. The vertical dashed lines indicate $0.12\,h\,{\rm Mpc}^{-1}$.}
    \label{fig:xpk}
\end{figure*}

\begin{figure}[htp]
    \centering
    \includegraphics[width=0.45\textwidth]{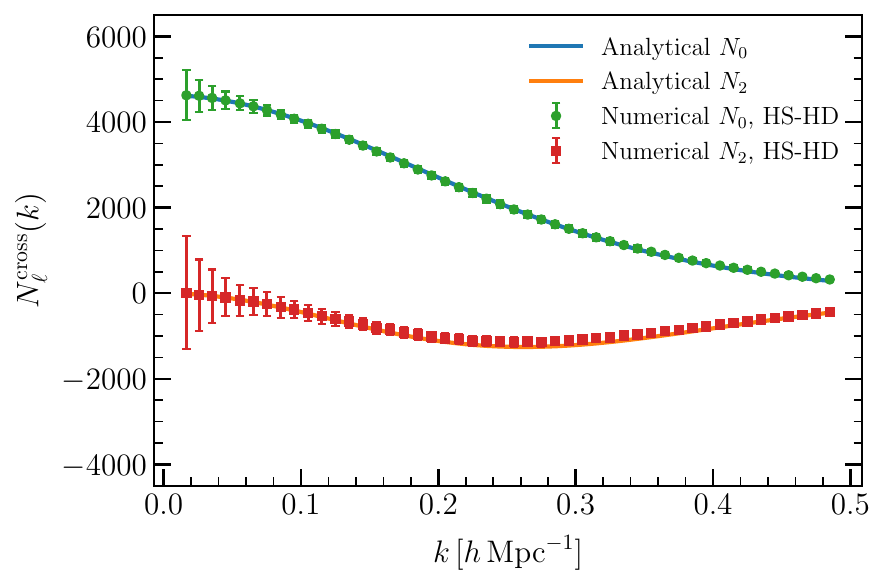} 
     \includegraphics[width=0.45\textwidth]{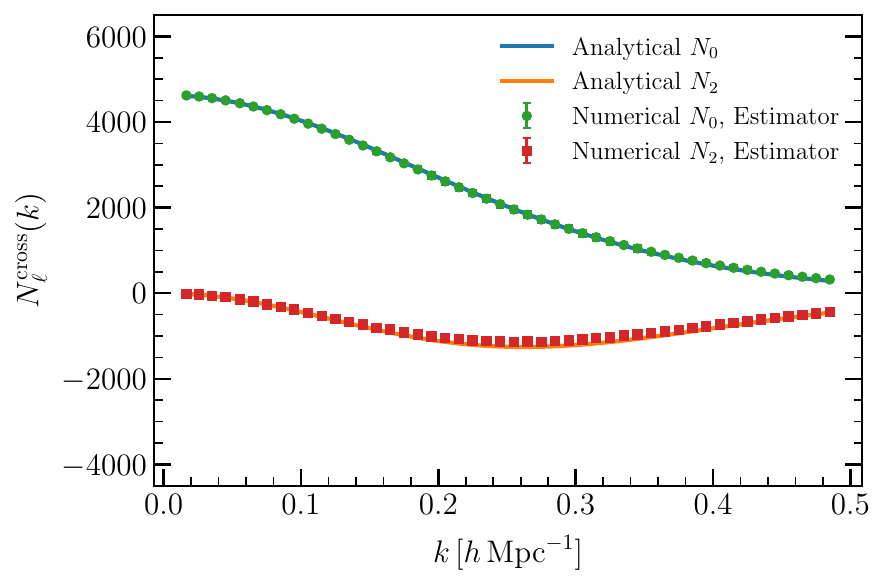} 
    \caption{Validation of the analytical model for the cross shot noise. The solid lines denote the theoretical prediction for $N_{\ell}^{\rm cross}$ obtained from Eqs.~(\ref{eq:Ncross2D}-\ref{eq:Ncrossell}) with $b=2.1, b_{\rm fid} = 1.824$, and $f=f_{\rm fid} = 0.778$. The data points show the measurement from GLAM mocks. The numerical results in the left and right panels are obtained using the HS-HD method in Eq.~(\ref{eq:xsn1}) and the new estimator in Eq.~(\ref{eq:xsn2}), respectively.}
    \label{fig:Nell}
\end{figure}

\section{Cross shot noise estimator}\label{sec:xSNestimator}

We use two approaches to estimate the shot noise of $P_{\rm cross}$ between the pre-reconstruction field $\delta$ and the post-reconstruction field $R$. For measurements in a periodic box, the cross-power-spectrum estimator including shot noise is given by,
\begin{align}
    \hat{P}^{\rm cross}_{\mathrm{w/,\,N}}(k, \mu) &= \left\langle \delta(\boldsymbol{k}) R(-\boldsymbol{k}) \right\rangle\,.
    \label{eq:xpkwSN}
\end{align}

\begin{itemize}
 \item \textbf{Method 1:} The cross power spectrum without shot noise can be estimated by the half-sum and half-difference (HS-HD) method \citep{Ando:2017wff, Wang:2022nlx}, where the galaxy catalog is randomly divided into two disjoint subsets, whose density fields are denoted by $\delta_1$ and $\delta_2$, with the corresponding reconstructed density fields denoted by $R_1$ and $R_2$, respectively,
 \begin{align}
    \hat{P}^{\rm cross}_{\mathrm{w/o,\,N}}(k, \mu) &= \left[\left\langle \delta_1(\boldsymbol{k}) R_2(-\boldsymbol{k}) \right\rangle +  \left\langle \delta_2(\boldsymbol{k}) R_1(-\boldsymbol{k}) \right\rangle \right]/2\,,
    \label{eq:xpk1}
\end{align}      
    
Taking the difference between cross power spectra with and without shot noise, the cross shot noise multipole is then obtained
 \begin{align}
    \hat N^{\rm cross}_{\ell}(k)= \frac{2 \ell+1}{2} \int d\mu {\cal L}_\ell(\mu) \left[ \hat{P}^{\rm cross}_{\mathrm{w/\,,N}}(k, \mu) -  \hat{P}^{\rm cross}_{\mathrm{w/o\,,N}}(k, \mu)\right] \,.
     \label{eq:xsn1}
\end{align}   

    \item \textbf{Method 2:} We propose a new estimator for the cross shot noise,  
\begin{equation}
    \mathrm{xSN}[\delta(-\boldsymbol{k})R(\boldsymbol{k})]=\sum_{i=1}^{N_{g}}e^{-i\boldsymbol{k}\cdot(\boldsymbol{x}_{{\rm rec}, i} - \boldsymbol{x}_{i})}\,.
\end{equation}
In practice, the summation is evaluated using Fast Fourier Transforms (FFT). We define a shifted noise catalog
\begin{equation}
n_{\rm xSN}(\boldsymbol{x})=\sum_{i}^{N_{g}} \delta_{D}(\boldsymbol{x}-(\boldsymbol{x}_{{\rm rec}, i} - \boldsymbol{x}_{i}))\,,
\end{equation}
and then paint it to mesh and perform Fourier transform. The cross shot noise multipole is then obtained,
\begin{equation}
    \hat N^{\rm cross}_{\ell}(k) =\frac{2\ell+1}{M_k} \frac{V}{N_g^2}  \sum_{\boldsymbol{k}}n_{\rm xSN}(\boldsymbol{k})\mathcal{L}_{\ell}(\hat{\boldsymbol{k}}\cdot\hat{\boldsymbol{z}})\,,
     \label{eq:xsn2}
\end{equation}
where $\sum_{\boldsymbol{k}}/M_k$ denotes the average over all Fourier modes $\boldsymbol{k}$ in a given $k$-bin. We perform the measurements using a modified version of \texttt{jax-power}\footnote{\url{https://github.com/adematti/jax-power/}}. Using the measured cross shot noise, we obtain the shot-noise-subtracted cross power spectrum multipoles as
 \begin{align}
   P^{\rm cross}_{\mathrm{\ell}}(k) &= \left[\frac{2 \ell+1}{2} \int d\mu {\cal L}_\ell(\mu) \hat{P}^{\rm cross}_{\mathrm{w/\,,N}}(k, \mu)\right] - \hat N^{\rm cross}_{\ell}(k) \,.
    \label{eq:xpk2}
\end{align}  
\end{itemize}

\section{Comparison with mocks} \label{sec:result}

To validate the Gaussian covariance matrix, we compare it with numerical covariance matrices estimated from the GLAM mock catalogues used in \citetjcap{Wang et al.}{Wang:2023hlx}. These mocks consist of $986$ independent realizations, with each realization covering a cubic volume of side length $1\,h^{-1}{\rm Gpc}$ \citep{Klypin:2017iwu}. We populate the simulations using the best-fit HOD parameters calibrated for the CMASS-like sample with $M_i<-21.6$ \citep{Guo:2014oca}, yielding a mean number density of $\bar n\simeq 2\times10^{-4}\,h^3{\rm Mpc}^{-3}$. 

We first compare the shot-noise-subtracted cross-power spectra and their variances obtained using the new estimator and the HS-HD method in Fig.~\ref{fig:xpk}. The mean measurements from the two methods agree almost perfectly. The variances agree well on large scales but gradually deviate with increasing $k$, reaching a difference of approximately 10\% at $k=0.25\,h\,{\rm Mpc}^{-1}$. The new estimator has lower variance at small scales as it does not require splitting catalogs. 

We examine the modelling of the cross shot noise given in Eqs.~(\ref{eq:Ncross2D}-\ref{eq:Ncrossell}). Fig.~\ref{fig:Nell} compares the analytical model with the measured cross shot-noise multipoles obtained using the HS-HD-based method in Eq.~(\ref{eq:xsn1}) and the new estimator in Eq.~(\ref{eq:xsn2}), shown in the left and right panels, respectively. The large uncertainties on large scales in the left panel arise from the propagation of the cosmic variance of the power-spectrum measurements. Both the measurements agree well the analytical model. The agreement demonstrates that the Gaussian displacement model provides an accurate description of the reconstruction-induced suppression of the cross shot noise.

\begin{figure*}[htp]
    \centering
    \includegraphics[width=0.99\textwidth]{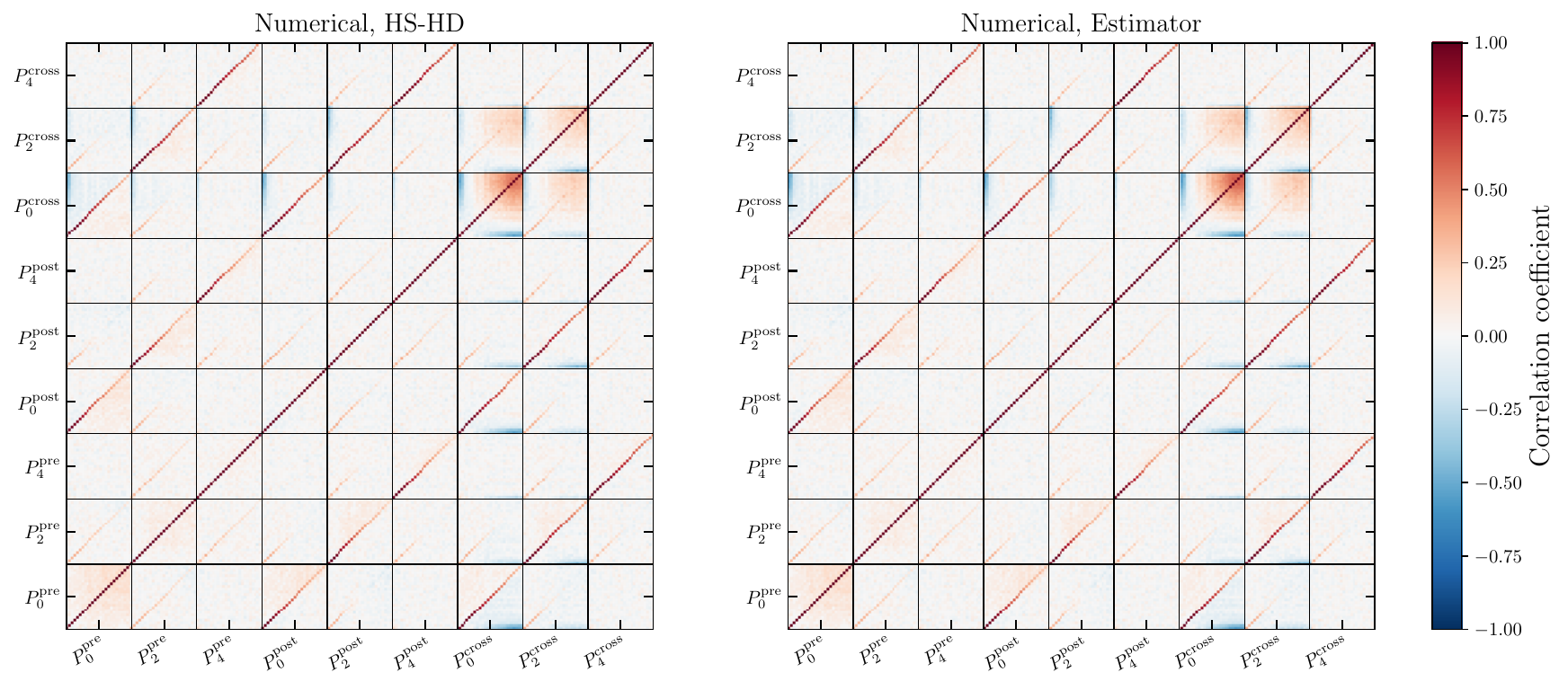}
    \caption{Correlation matrices estimated from the mock catalogues using the HS-HD method in Eq.~(\ref{eq:xpk1}) (left) and the new estimator in Eq.~(\ref{eq:xpk2}) (right) over the range $0<k<0.25\,h\,{\rm Mpc}^{-1}$. The covariance includes the auto- and cross-correlations among the pre-reconstruction, post-reconstruction, and cross power-spectrum multipoles $(\ell=0,2,4)$.}
    \label{fig:corrmatNum}
\end{figure*}

\begin{figure*}[htp]
    \centering
    \includegraphics[width=0.5\textwidth]{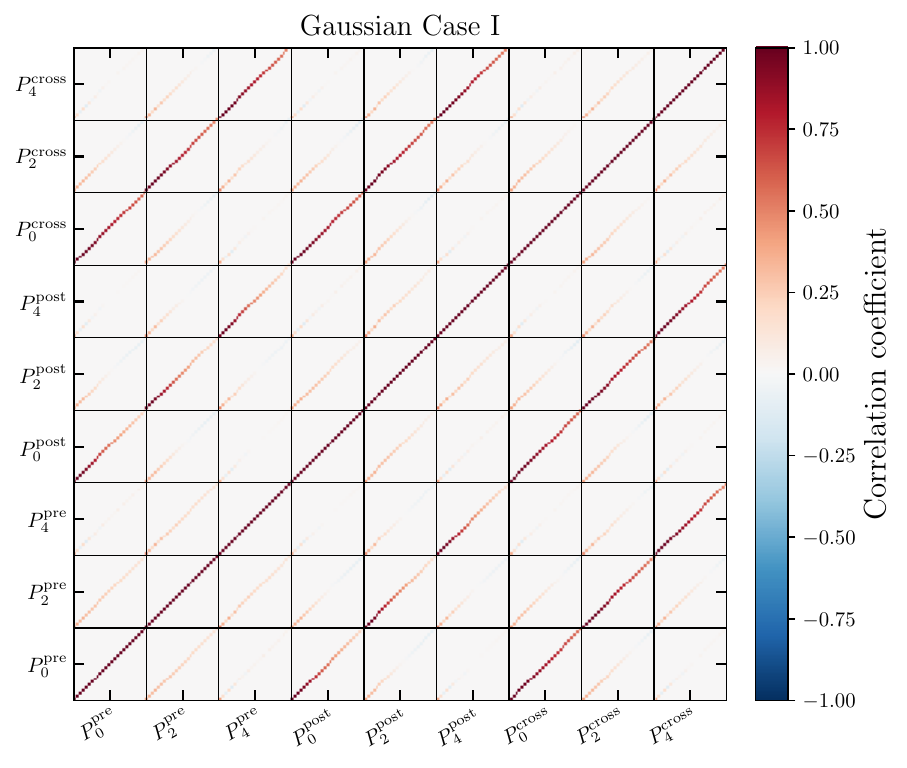}
    \caption{Correlation matrix predicted by the semi-analytical \textbf{Gaussian case I}. }
    \label{fig:corrmatGau}
\end{figure*}

\begin{figure*}[htp]
    \centering
    \includegraphics[width=0.9\textwidth]{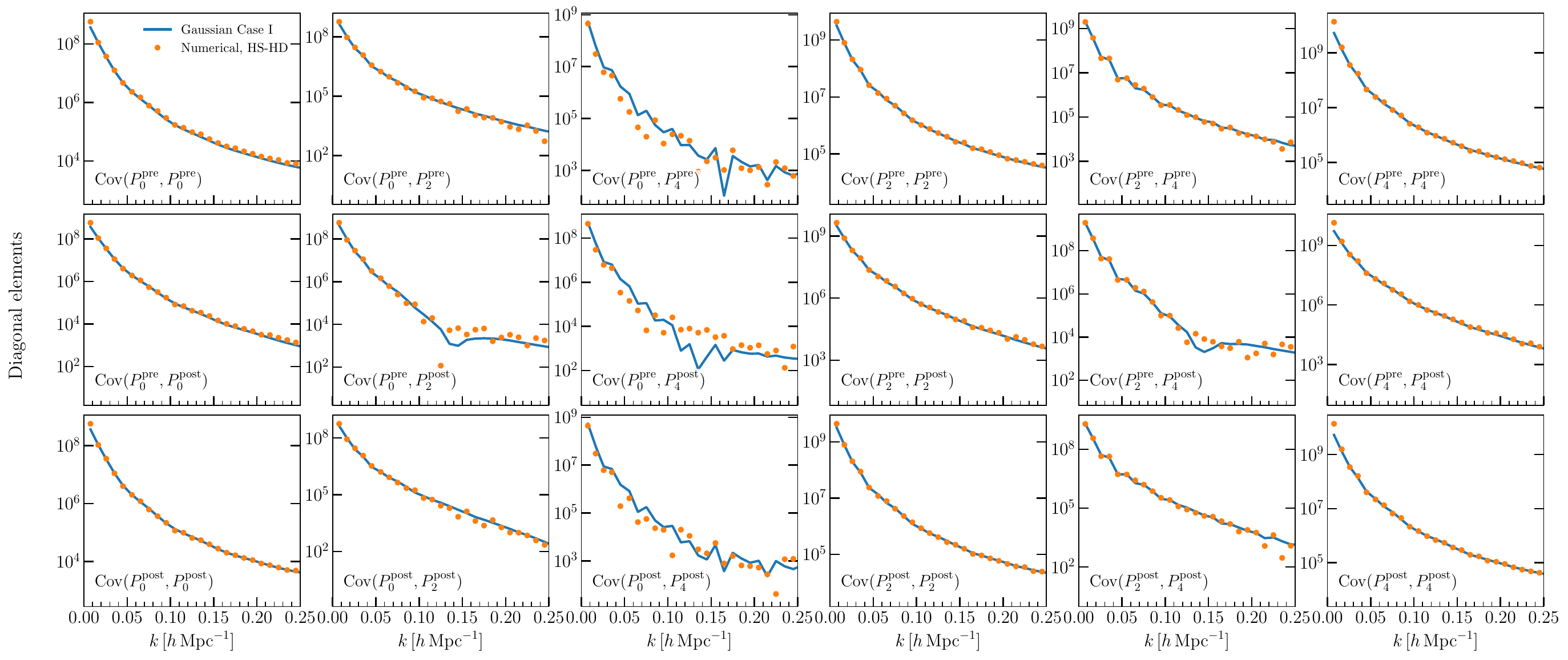}
    \caption{Diagonal elements of the covariance matrix for $P_{\ell}^{\rm pre}$ (top row), the cross-covariance between between $P_{\ell}^{\rm pre}$ and $P_{\ell}^{\rm post}$ (middle row), and the covariance for $P_{\ell}^{\rm post}$ (bottom row). In each panel, the solid line denotes the semi-analytical Gaussian prediction, while the symbols show the numerical measurements from mocks. The choice of the $P_{\rm cross}$ measurement only affects covariance terms involving $P_{\ell}^{\rm cross}$ and does not change the covariance between $P_{\ell}^{\rm pre}$ and $P_{\ell}^{\rm post}$. Therefore, for the $P_{\ell}^{\rm pre}$ and $P_{\ell}^{\rm post}$ covariance block, we show only the \textbf{Gaussian Case I} prediction and the HS-HD-based measurement.}
    \label{fig:preell}
\end{figure*}

Using the mean power-spectrum multipoles measured from the GLAM mock catalogues as inputs, we construct the semi-analytical covariance matrix following the Gaussian covariance formalism described in Sec.~\ref{sec:covmat}. The shot noise contributions are included using the models presented in Sec.~\ref{sec:xshotnoise}. This allows us to predict the full Gaussian covariance matrix of the joint data vector consisting of the pre-reconstruction, post-reconstruction, and cross power-spectrum multipoles. Since $P_{\rm cross}$ can be measured in different ways, we consider three implementations of the semi-analytical Gaussian covariance,
\begin{itemize}
    \item \textbf{Gaussian Case I}: HS-HD-based cross power spectrum (Eq.~\ref{eq:xpk1}) with the analytical cross shot-noise model (Eqs.~\ref{eq:Ncross2D}-\ref{eq:Ncrossell});
    \item \textbf{Gaussian Case II}: Estimator-based cross power spectrum (Eq.~\ref{eq:xpk2}) with the analytical cross shot-noise model(Eqs.~\ref{eq:Ncross2D}-\ref{eq:Ncrossell}) ;
    \item \textbf{Gaussian Case III}: Cross power spectrum measured directly with shot noise included (Eq.~\ref{eq:xpkwSN}) .
\end{itemize}

\begin{figure*}[htp]
    \centering
    \includegraphics[width=0.9\textwidth]{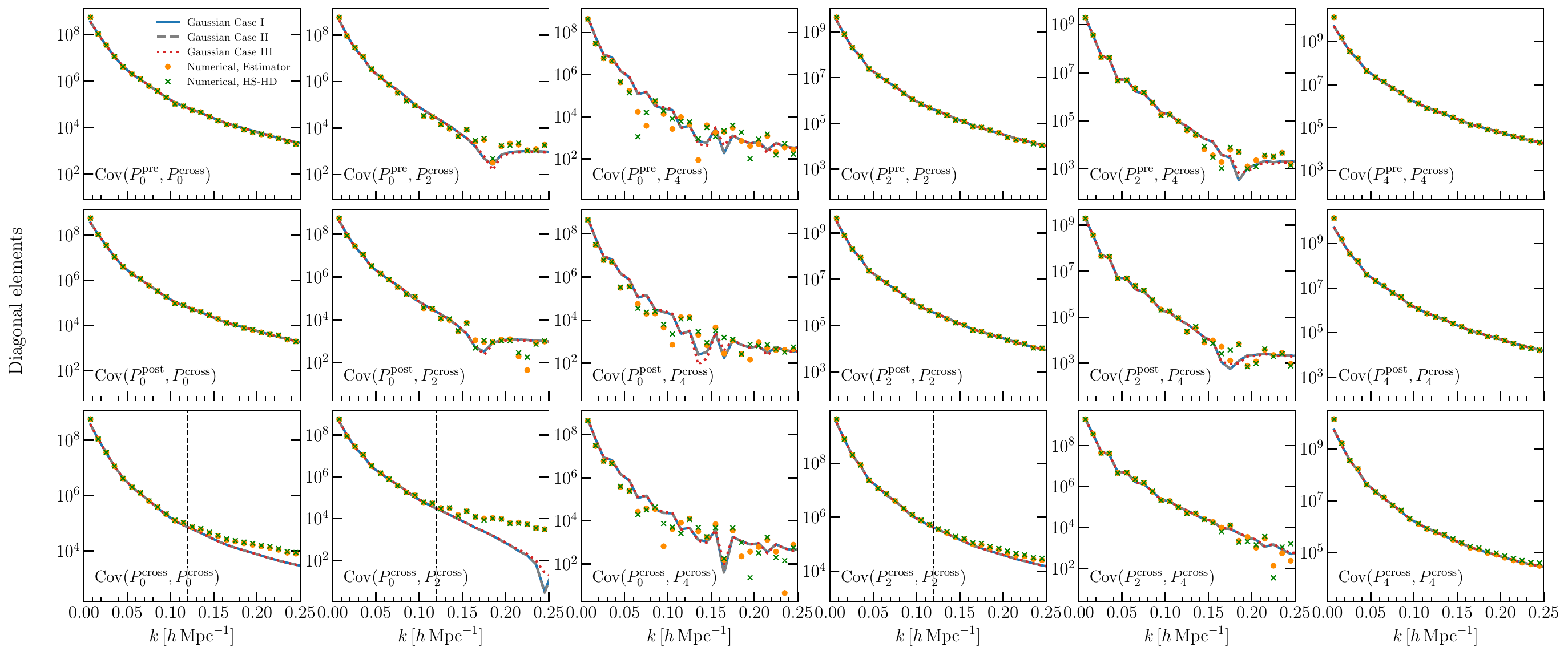}
    \caption{Same as Fig.~\ref{fig:preell}, but for the diagonal elements of the cross covariance between $P_{\ell}^{\rm pre}$ and $P_{\ell}^{\rm cross}$ (top row), the cross covariance between $P_{\ell}^{\rm post}$ and $P_{\ell}^{\rm cross}$ (middle row), and the auto covariance of $P_{\ell}^{\rm cross}$ (bottom row). In each panel, the solid, dashed, and dotted lines denote the semi-analytical Gaussian predictions in three cases, while the symbols show the numerical measurements from mocks: circles correspond to the new cross shot-noise estimator, and crosses to the HS-HD-based results. The dashed vertical lines in the bottom row mark the scale range $k= 0.12\,h\,{\rm Mpc}^{-1}$ for the covariance blocks involving $P_0$ and $P_2$, over which the semi-analytical Gaussian covariance and numerical covariances show good agreement.}
    \label{fig:crossell}
\end{figure*}

Fig.~\ref{fig:corrmatNum} presents the correlation matrices measured from the GLAM mock catalogues using the two numerical methods, while Fig.~\ref{fig:corrmatGau} shows an example prediction from the semi-analytical \textbf{Gaussian case I}. Since the three Gaussian predictions only affect covariance terms involving $P_{\rm cross}$, the correlation matrix structure remains unchanged for the three Gaussian cases. The comparison of diagonal covariance parts among the three Gaussian predictions is presented in Fig.~\ref{fig:crossell}. The correlation matrix includes all auto- and cross-correlations among the pre-reconstruction, post-reconstruction, and cross power-spectrum multipoles with $\ell=0,2,4$. It is seen that the semi-analytical Gaussian covariance model mainly reproduces the diagonal structure of individual covariance blocks, including the auto-correlations of each observable and the cross-correlations among different power-spectrum multipoles. 

We then compare the diagonal elements of the covariance matrix in different covariance blocks, as shown in Figs.~\ref{fig:preell} and \ref{fig:crossell}. For the pre-reconstruction power spectrum multipoles, the semi-analytical Gaussian covariance accurately describes both the amplitude and scale dependence of the diagonal covariance elements involving $P_0$ and $P_2$ up to $k=0.25\,h\,{\rm Mpc}^{-1}$ (top row in Fig.~\ref{fig:preell}). Similar agreement is found for the post-reconstruction multipoles (bottom row in Fig.~\ref{fig:preell}), and for the covariance between $P_{\ell}^{\rm pre}$ and $P_{\ell}^{\rm post}$ ($\ell=0,2$), with deviations for ${\rm Cov}(P_0^{\rm pre}, P_2^{\rm post})$ appearing at higher $k$. For the covariance blocks involving $P_4$, the semi-analytical Gaussian covariance generally follows the numerical covariance, although the mixed $P_0$-$P_4$ covariance block exhibits noticeable fluctuations.

For the $P_{\ell}^{\rm cross}$-related covariance terms, we show the three semi-analytical Gaussian predictions together with the two numerical measurements in Fig. \ref{fig:crossell}. The three Gaussian covariance predictions are nearly identical, indicating that the choice of $P_{\ell}^{\rm cross}$ and its shot-noise treatment has a negligible impact on the Gaussian covariance. The semi-analytical Gaussian covariance accurately captures the diagonal covariance elements between $P_{\ell}^{\rm pre/post}$ and $P_{\ell}^{\rm cross}$ involving $P_0$ and $P_2$, with deviations mainly appearing in the $P_0$-$P_2$ cross-correlations at higher $k$. For the auto-covariance of $P_{\ell}^{\rm cross}$, shown in the bottom row of Fig.~\ref{fig:crossell}, the Gaussian predictions agrees well with the numerical covariance for $P_0$ and $P_2$ up to $k=0.12\,h\,{\rm Mpc}^{-1}$. At higher $k$, small deviations appear. These differences are expected, as nonlinear mode coupling and non-Gaussian contributions become increasingly important on smaller scales for $P_{\rm cross}$. For the covariance blocks involving $P_4$, the overall agreement remains reasonable, although noticeable fluctuations are seen for the mixed $P_0$-$P_4$ covariance block.

\begin{figure*}[htp]
    \centering
    \includegraphics[width=0.3\textwidth]{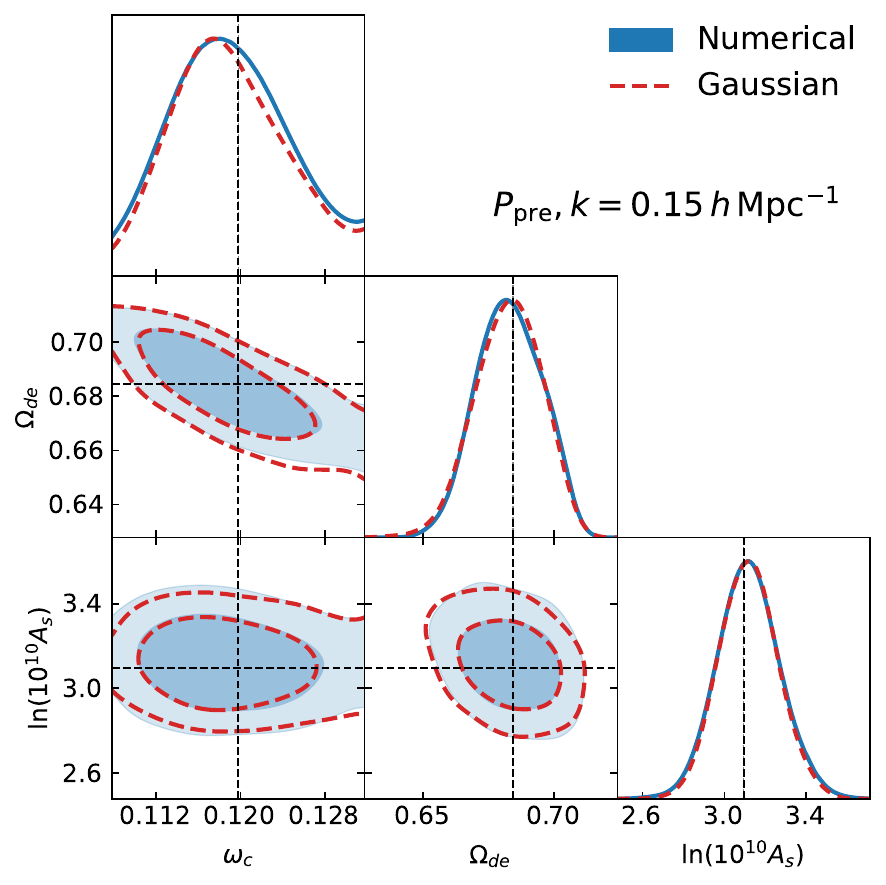}
    \includegraphics[width=0.3\textwidth]{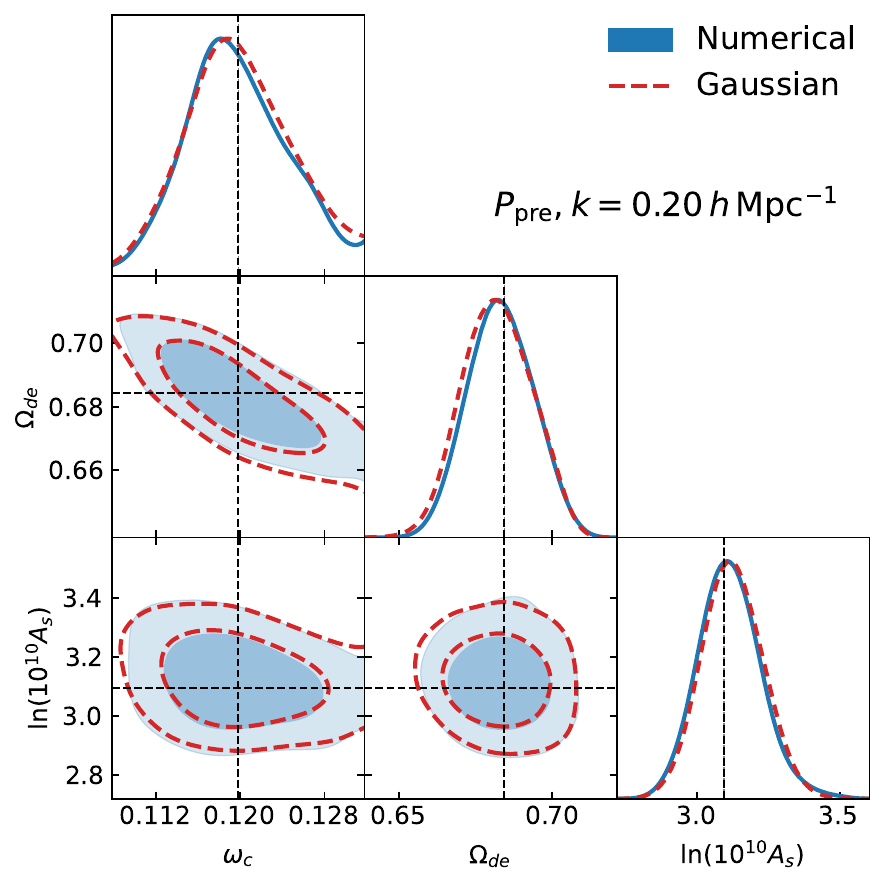}
    \includegraphics[width=0.3\textwidth]{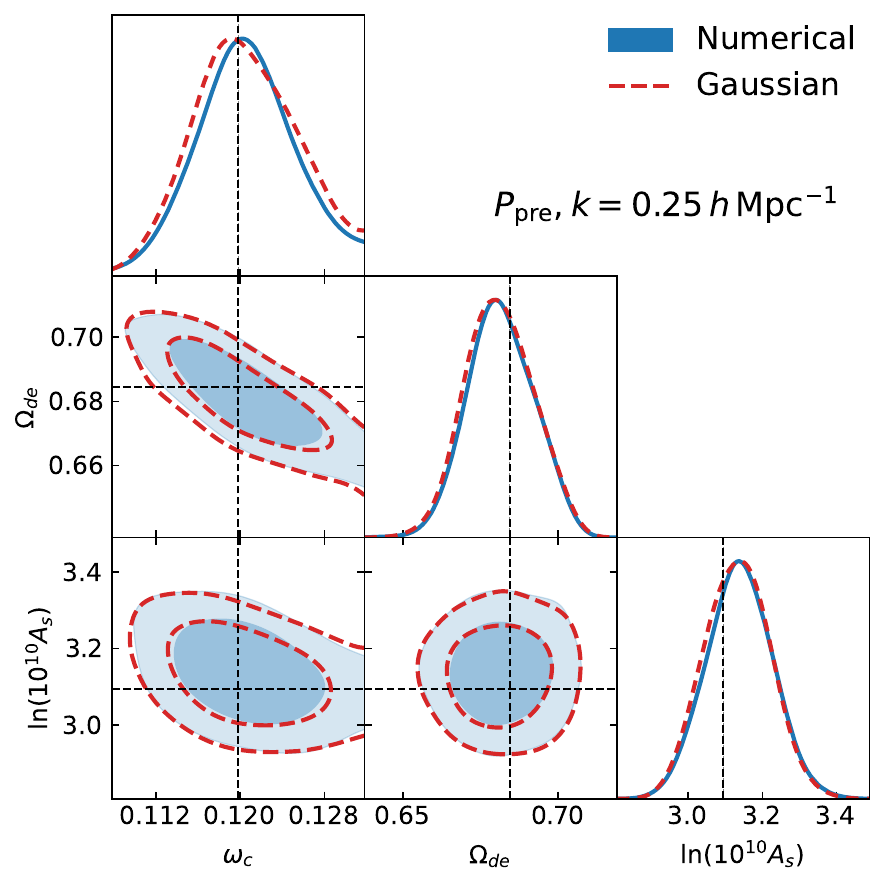}
    \caption{Comparison of parameter constraints from the pre-reconstruction power spectrum multipoles $(P^{\rm pre}_{\ell})$ obtained using the Gaussian covariance matrix and the numerical covariance estimated from the GLAM mock catalogs. The results are shown for three different maximum wavenumbers.}
    \label{fig:preonly}
\end{figure*}

\begin{figure*}[!htbp]
    \centering
    \includegraphics[width=0.3\textwidth]{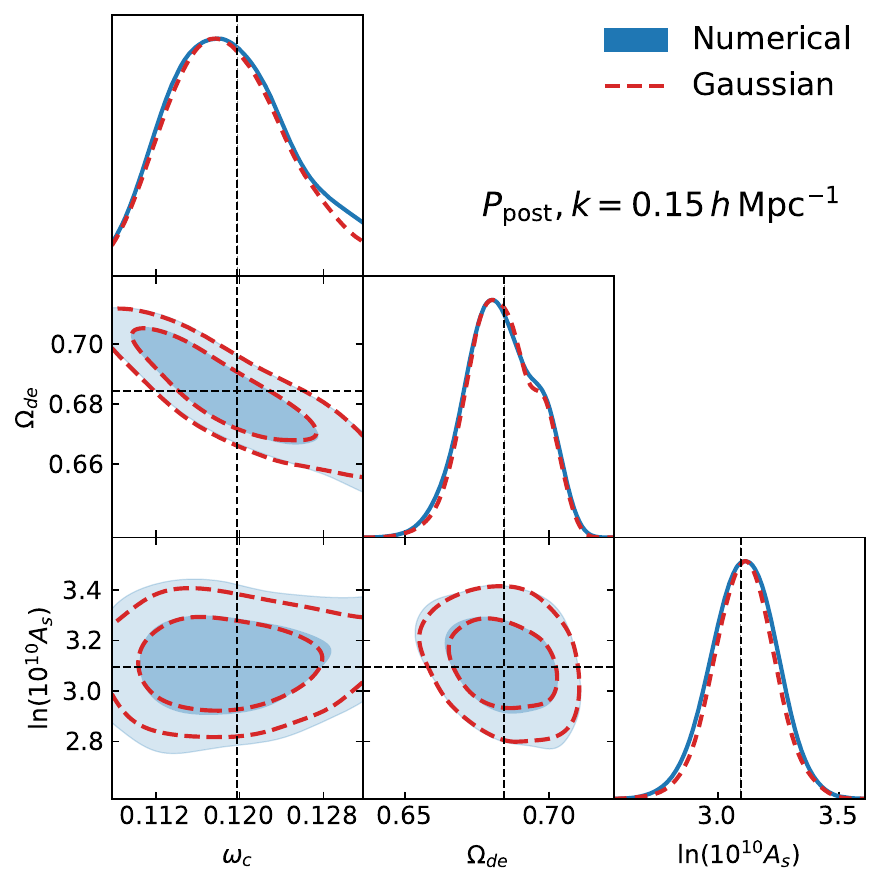}
    \includegraphics[width=0.3\textwidth]{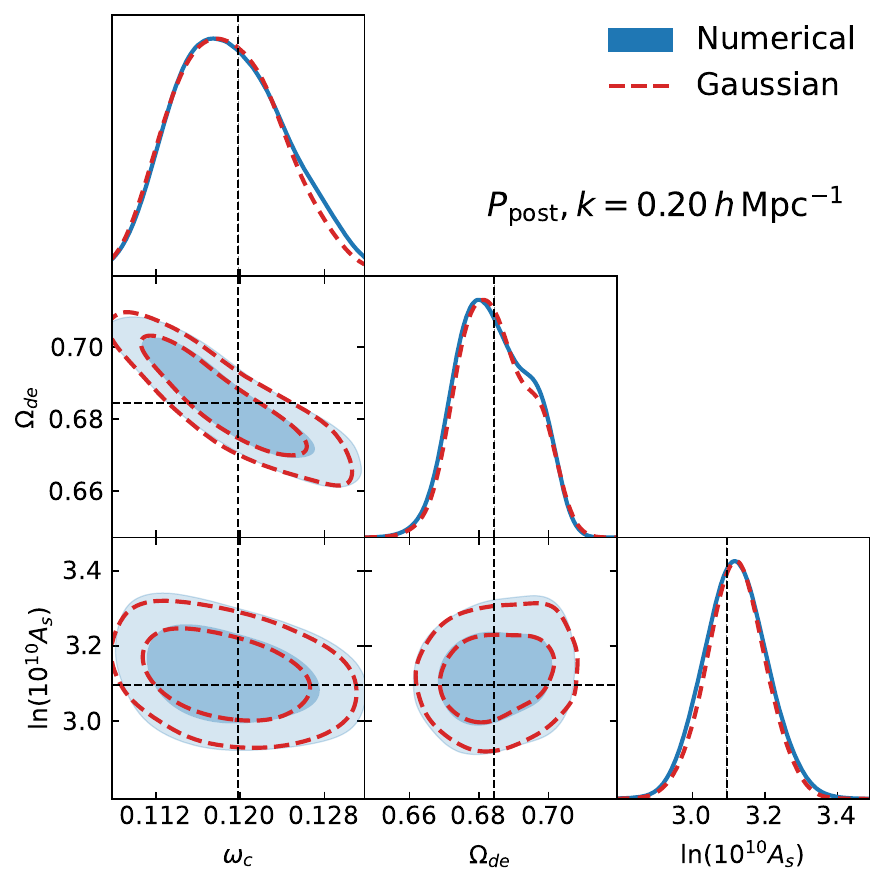}
        \includegraphics[width=0.3\textwidth]{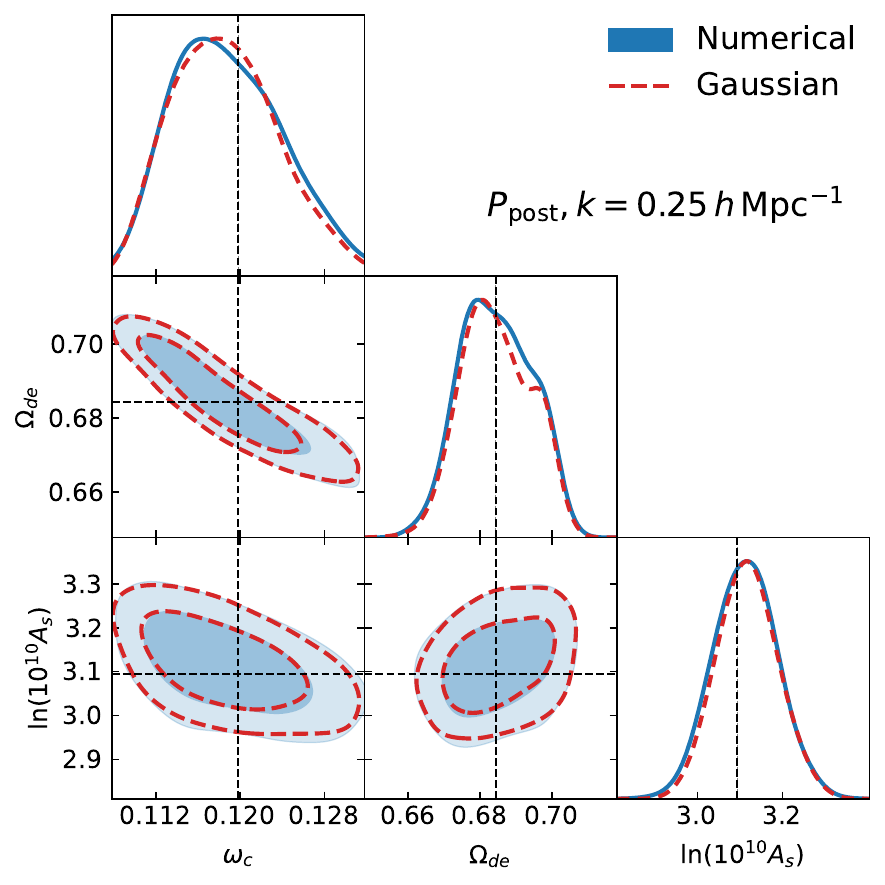}
    \caption{Same as Fig.~\ref{fig:preonly}, but from the post-reconstruction power spectrum multipoles $P^{\rm post}_{\ell}$.}
    \label{fig:postonly}
\end{figure*}

\begin{figure*}[!htbp]
    \centering
    \includegraphics[width=0.3\textwidth]{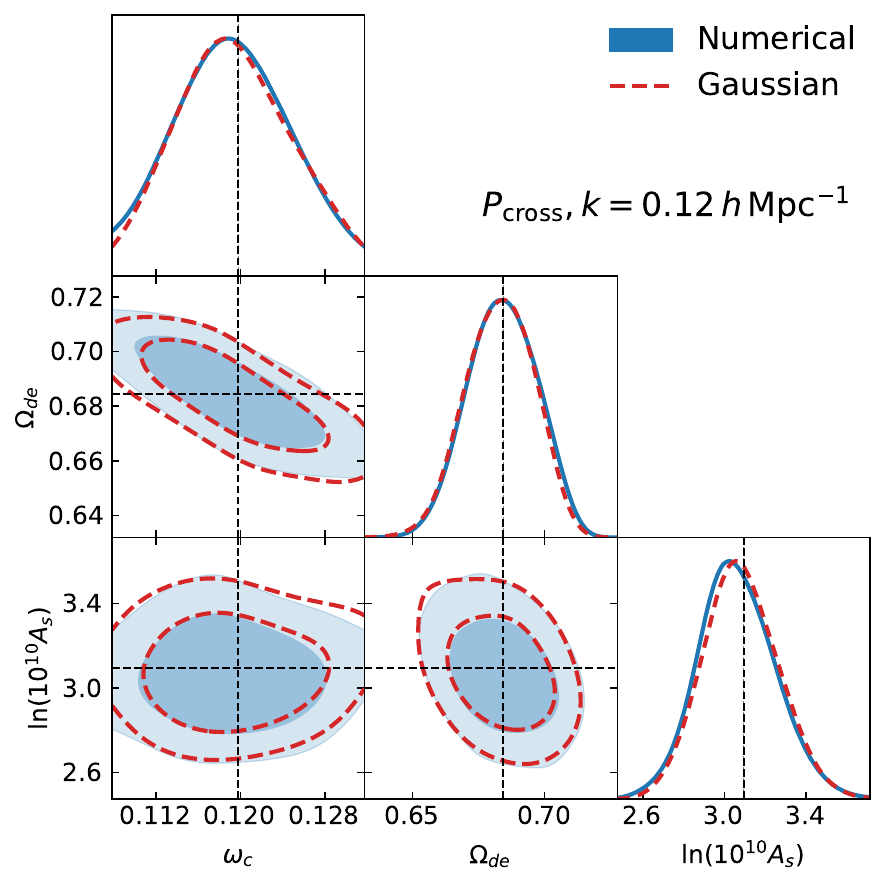}
    \includegraphics[width=0.3\textwidth]{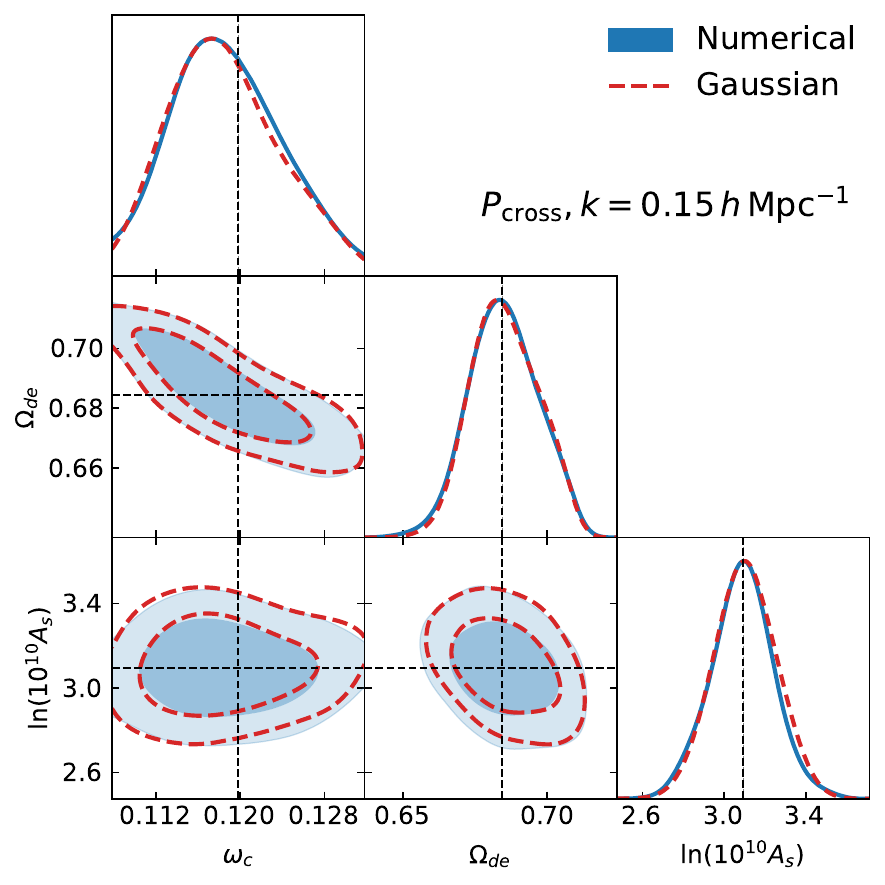}
    \includegraphics[width=0.3\textwidth]{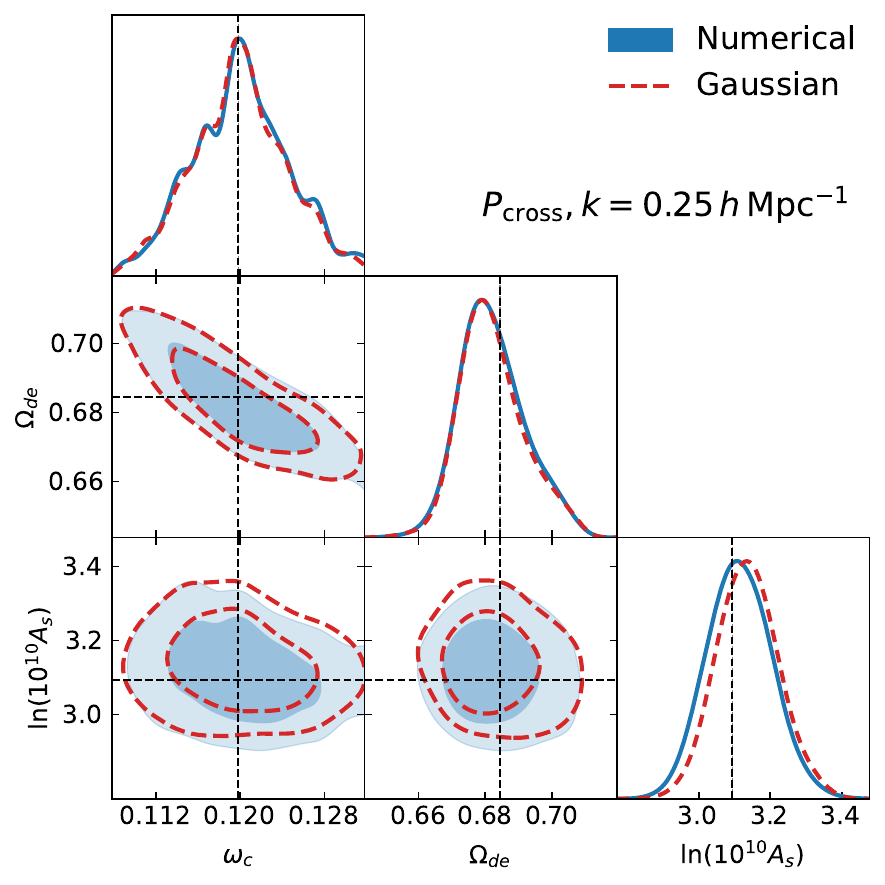}
    \caption{Same as Fig.~\ref{fig:preonly}, but from the cross power spectrum multipoles $P^{\rm cross}_{\ell}$.}
    \label{fig:xonly}
\end{figure*}

To further validate the impact of the covariance matrix on cosmological parameter inference, we perform parameter fits using both the semi-analytical Gaussian covariance and the numerical covariance estimated from the GLAM mocks. We adopt the fiducial cosmology prediction from the emulator \citep{Wang:2023hlx} as the mock observation vector. The theoretical model is evaluated using the same emulator, with the emulator uncertainty incorporated into the statistical covariance matrix \citep{Wang:2023hlx}. Since the emulator for $P_{\rm cross}$ in \cite{Wang:2023hlx} is based on the HS-HD method, we adopt the numerical covariance estimated using the HS-HD-based measurements in the fitting analysis. As the three implementations of the semi-analytical Gaussian covariance are indistinguishable. Therefore, we use \textbf{Gaussian Case I} as the Gaussian covariance in all subsequent likelihood analyses. For the numerical covariance, we apply the \citetjcap{Percival}{Percival:2013sga} correction to the parameter covariance to account for the finite number of mock realizations. In the parameter fit, we use only the monopole and quadrupole power spectra ($\ell=0$ and $2$). The hexadecapole is not included because its emulator uncertainty is relatively large and its inclusion provides little additional constraining power beyond $P_0$ and $P_2$ \citep{Wang:2023hlx}. Both the semi-analytical Gaussian covariance and the numerical covariance matrices are rescaled to correspond to an effective survey volume three times that of the GLAM mock, i.e., $3 (\,h^{-1}\,{\rm Gpc})^{3}$.

For the pre-reconstruction power spectrum, we perform parameter fits with $k_{\rm max}=0.15$, $0.20$, and $0.25\,h\,{\rm Mpc}^{-1}$, corresponding to Percival correction factors of $\mathcal{P} = 1.044$, $1.066$, and $1.089$, respectively. The constraints obtained with the Gaussian and numerical covariance matrices are in good agreement over the full range of $k_{\rm max}$, as shown in Fig.~\ref{fig:preonly}, indicating that the off-diagonal covariance contributes only weakly to the parameter uncertainties. The same conclusion also holds for the post-reconstruction power spectrum, as shown in Fig.~\ref{fig:postonly}.

For the cross power spectrum, we consider $k_{\rm max}=0.12$, $0.15$, and $0.25\,h\,{\rm Mpc}^{-1}$, shown in Fig.~\ref{fig:xonly}, corresponding to Percival correction factors of $\mathcal{P} = 1.031$, $1.044$, and $1.089$, respectively. At $k_{\rm max}=0.12, 0.15\,h\,{\rm Mpc}^{-1}$, the constraints obtained with the two covariance matrices remain marginally consistent. As $k_{\rm max}$ increases to $0.25\,h\,{\rm Mpc}^{-1}$, the difference begin to emerge. This behavior indicates that the Gaussian covariance approximation becomes progressively less accurate for the cross power spectrum, implying that the non-Gaussian contribution, particularly the off-diagonal covariance, becomes increasingly important on smaller scales. 

\begin{figure*}[htp]
    \centering
    \includegraphics[width=0.3\textwidth]{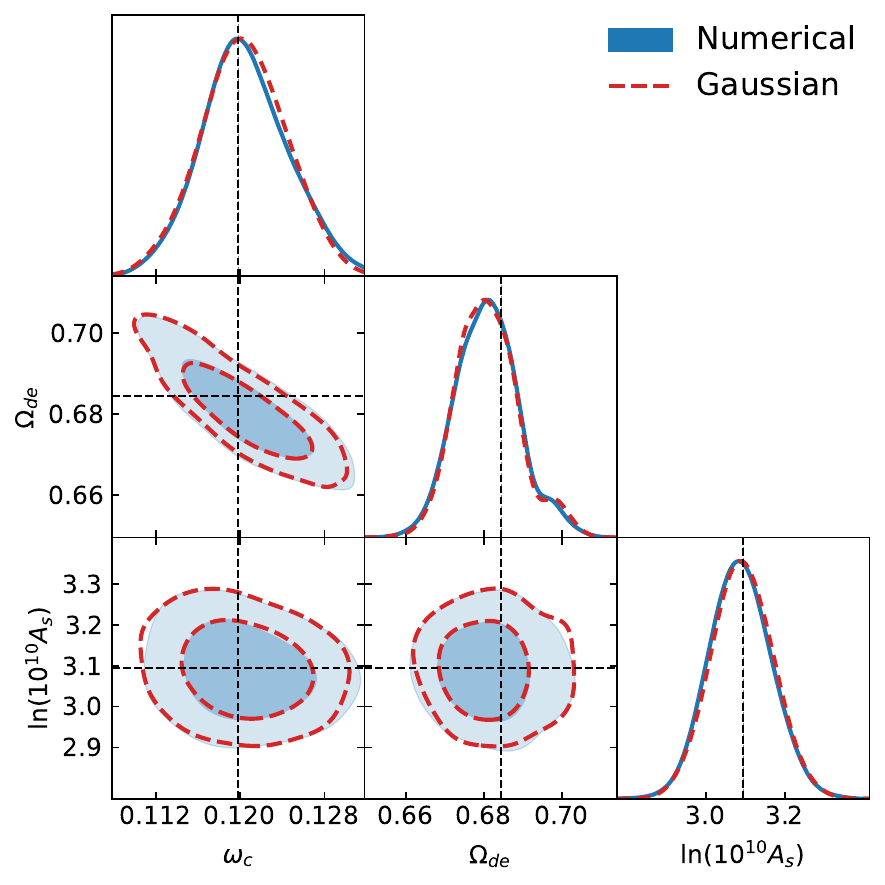}
    \includegraphics[width=0.3\textwidth]{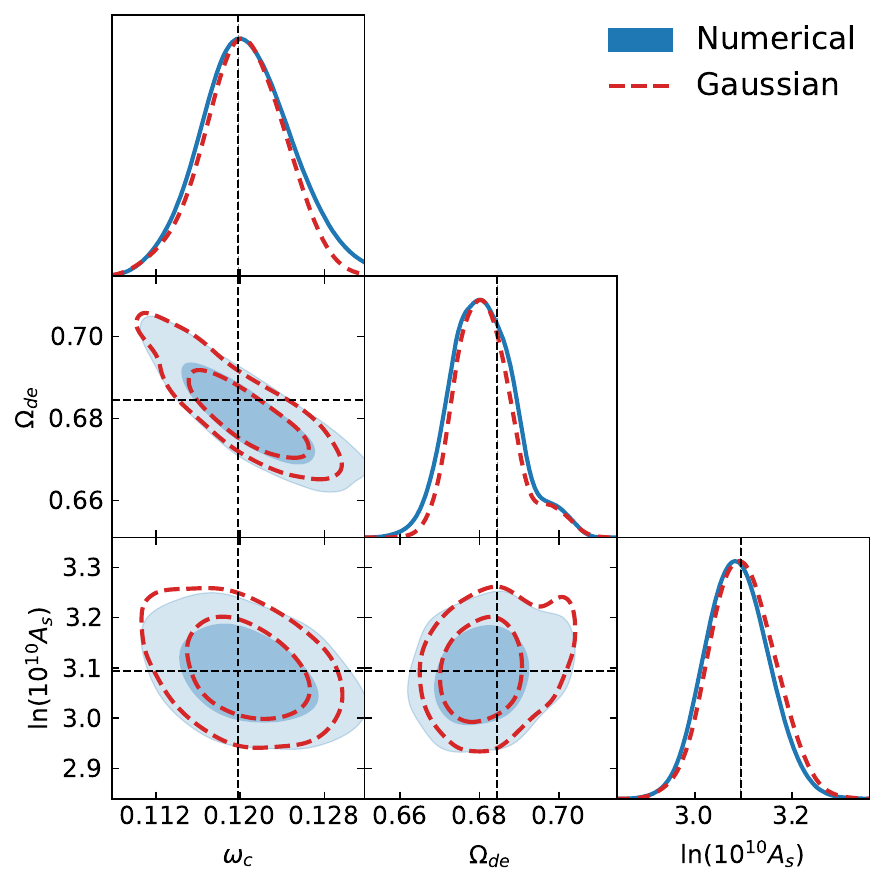}
     \includegraphics[width=0.3\textwidth]{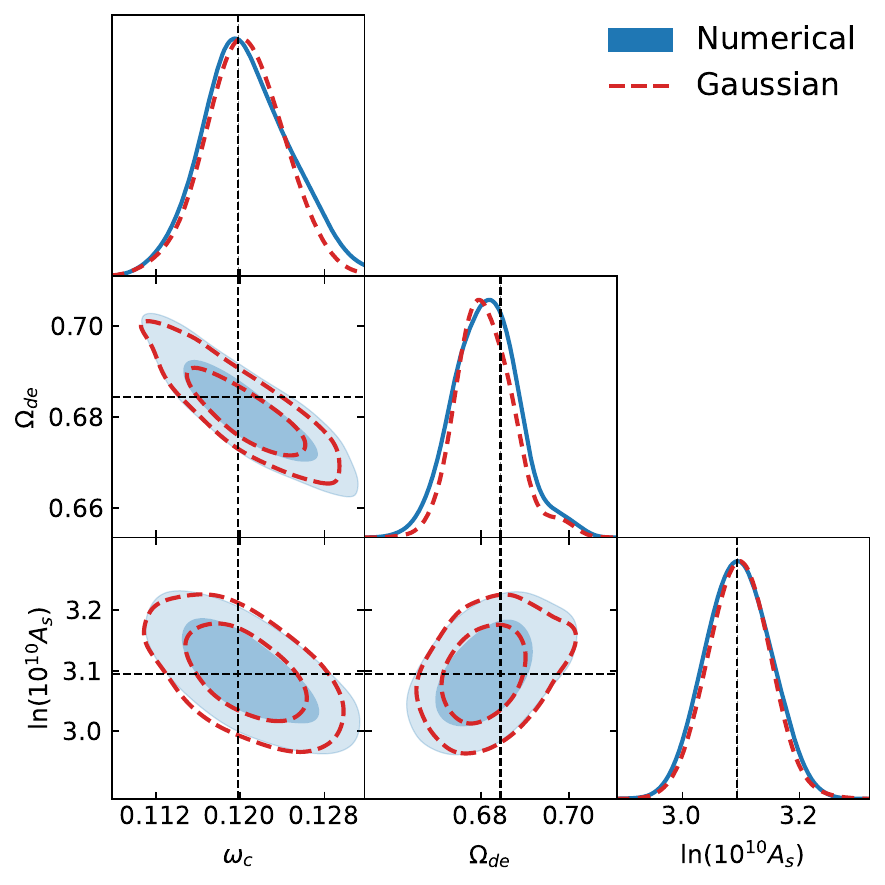}
    \caption{Comparison of cosmological parameter constraints from the joint analysis of $P_{\rm pre} + P_{\rm post} + P_{\rm cross}$ obtained using the semi-analytical Gaussian covariance matrix and the numerical covariance estimated from the GLAM mock catalogs. The three panels correspond to different choices of the maximum fitting scale. Left: $k_{\rm max}=0.18\,h\,{\rm Mpc}^{-1}$ for $P_{\rm pre}$ and $P_{\rm post}$, and $k_{\rm max}=0.12\,h\,{\rm Mpc}^{-1}$ for $P_{\rm cross}$. The Percival correction factor for numerical covariance is $\mathcal{P}= 1.204$. Middle: $k_{\rm max}=0.2 \,h\,{\rm Mpc}^{-1}$ for $P_{\rm pre}$ and $P_{\rm post}$, and $k_{\rm max}=0.15\,h\,{\rm Mpc}^{-1}$ for $P_{\rm cross}$. $\mathcal{P}= 1.243$. Right: $k_{\rm max}=0.25 \,h\,{\rm Mpc}^{-1}$ for $P_{\rm pre}$ and $P_{\rm post}$, and $k_{\rm max}=0.15 \,h\,{\rm Mpc}^{-1}$ for $P_{\rm cross}$. $\mathcal{P}= 1.302$.}
    \label{fig:alldiffk}
\end{figure*}

\begin{figure*}[htp]
    \centering
    \includegraphics[width=0.45\textwidth]{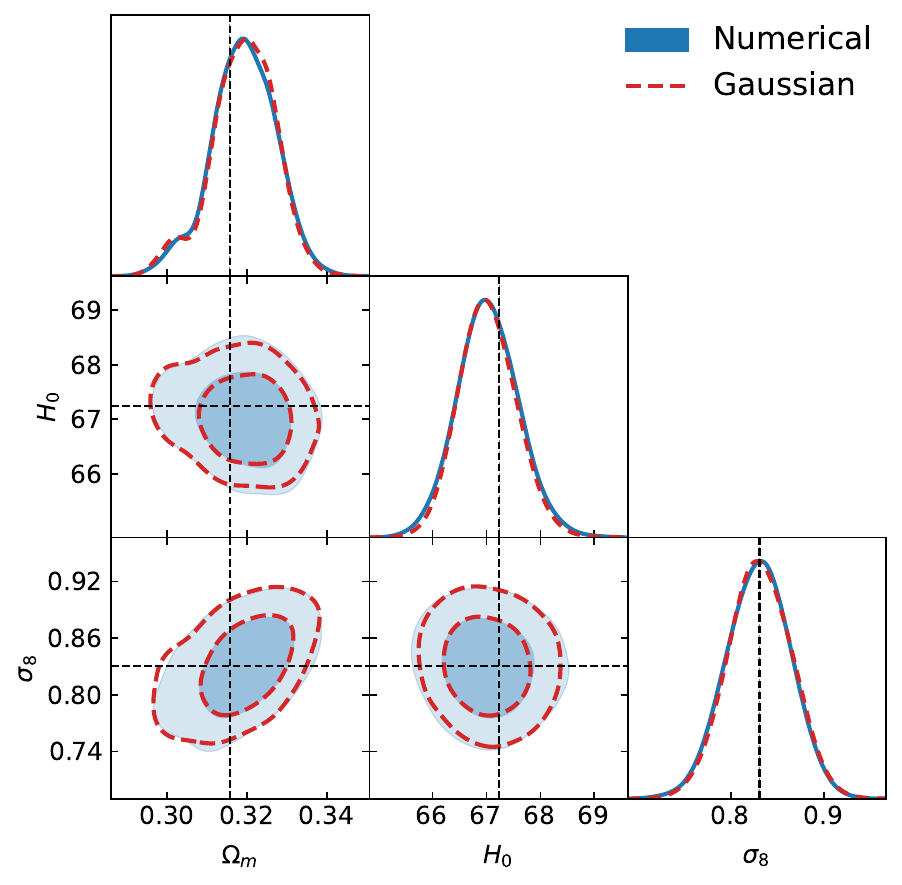}
    \includegraphics[width=0.45\textwidth]{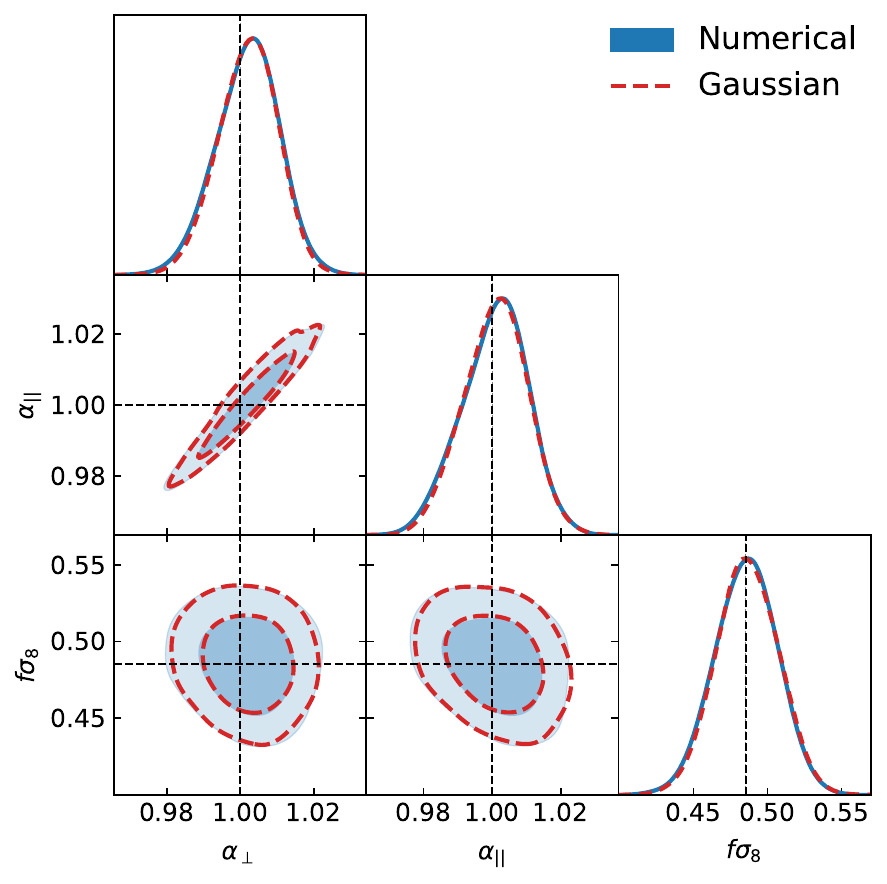}
    \caption{Comparison of derived parameter constraints from the joint analysis of $P_{\rm pre} + P_{\rm post} + P_{\rm cross}$ using the Gaussian covariance matrix and the numerical covariance estimated from the GLAM mock catalogs. The left panel shows the constraints on ($\Omega_m, H_0, \sigma_8$), while the right panel shows the constraints on ($\alpha_{\perp}, \alpha_{||}, f\sigma_8$). For the pre- and post-reconstruction power spectra, modes up to $k_{\rm max}=0.18\,h\,{\rm Mpc}^{-1}$ are included, whereas for the cross-power spectrum modes with $k_{\rm max}=0.12\,h\,{\rm Mpc}^{-1}$ are used.}
    \label{fig:fit}
\end{figure*}

Then we consider the joint analysis of the pre-reconstruction, post-reconstruction, and cross power-spectrum multipoles. Fig.~\ref{fig:alldiffk} compares the cosmological parameter constraints from the joint analysis of $P_{\rm pre} + P_{\rm post} + P_{\rm cross}$ obtained using the semi-analytical Gaussian covariance matrix and the numerical covariance estimated from the GLAM mock catalogues for different fitting ranges. For the fitting range, with $k_{\rm max}=0.18\,h\,{\rm Mpc}^{-1}$ for $P_{\rm pre}$ and $P_{\rm post}$, and $k_{\rm max}=0.12\,h\,{\rm Mpc}^{-1}$ for $P_{\rm cross}$, the two covariance matrices yield nearly identical cosmological parameter constraints in the left panel of Fig.~\ref{fig:alldiffk}. We also present the constraints on the derived cosmological parameters $(\Omega_m,H_0,\sigma_8)$, and BAO and RSD parameters $(\alpha_\perp,\alpha_{||}, f\sigma_8)$ in Fig.~\ref{fig:fit}, showing excellent agreement.

As the fitting range is extended to smaller scales, as shown in the middle and right panels of Fig.~\ref{fig:alldiffk}, corresponding to larger $\mathcal{P}$ correction factors, the differences begin to emerge, indicating that non-Gaussian covariance becomes increasingly important in the joint analysis.

The consistency between the parameter constraints demonstrates that the semi-analytical Gaussian covariance matrix accurately captures the dominant contribution to covariance structure of the full data vector. This provides a computationally efficient alternative to covariance estimation from a large number of mock catalogues for joint full-shape analyses. Although the pre- and post-reconstruction power spectra can individually be extended to larger $k$ values, their joint analysis introduces a higher-dimensional data vector and therefore requires a larger $\mathcal{P}$ correction factor. In addition, the cross covariance between the three types of spectra also needs to be taken into account. The resulting parameter constraints obtained with the Gaussian and numerical covariance matrices exhibit some differences. We therefore adopt the more conservative $k$-range to ensure the robustness of the covariance validation. Moreover, for $P_{\rm cross}$ at $k \leq 0.12 \,h\,{\rm Mpc}^{-1}$, the variances estimated using the HS-HD method and the new estimator differ by less than 2\%, indicating that the impact of the estimator choice is subdominant over the adopted fitting range.

\begin{figure*}[htp]
    \centering
    \includegraphics[width=0.45\textwidth]{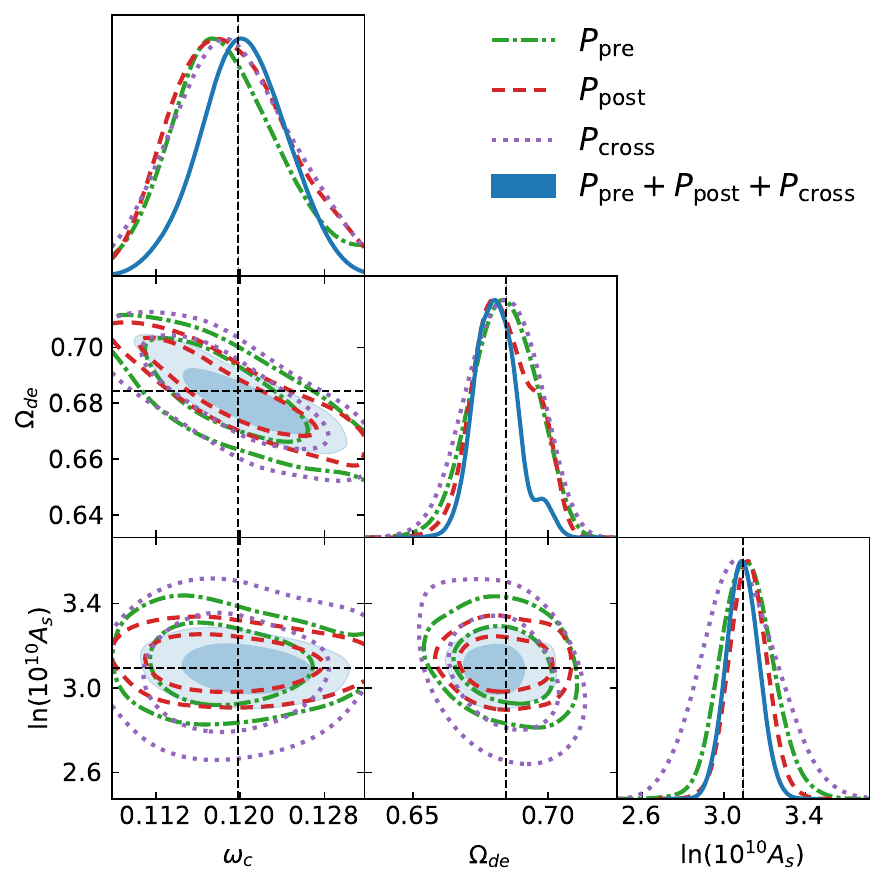}
    \includegraphics[width=0.45\textwidth]{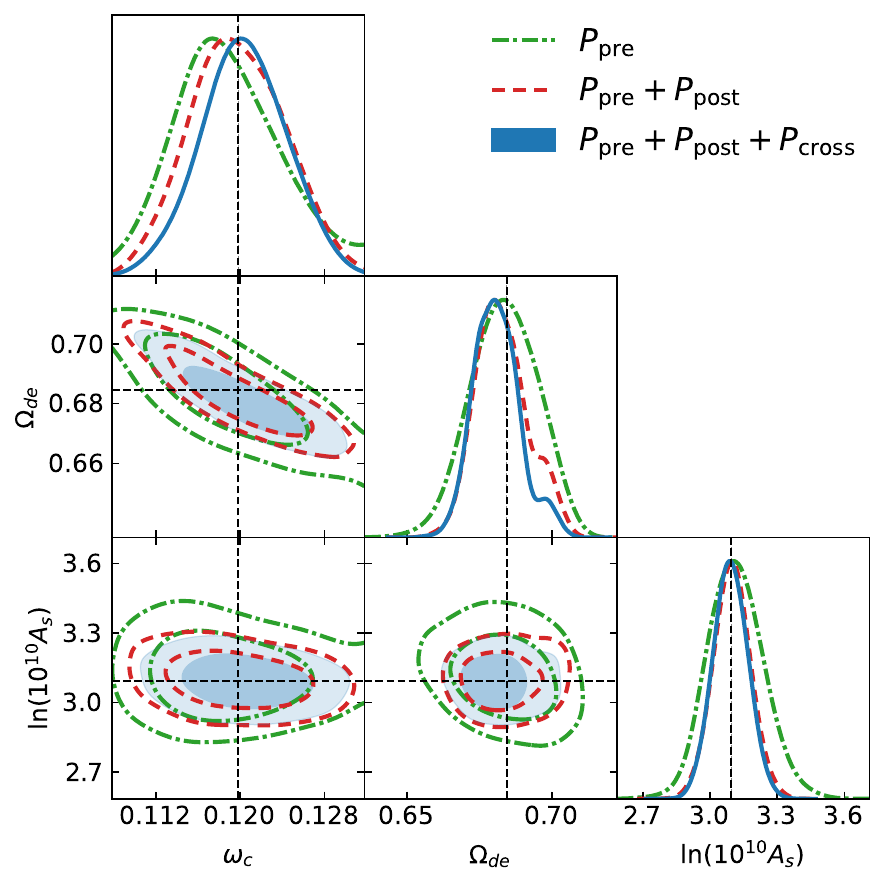}
    \caption{Cosmological parameter constraints from different combinations using Gaussian covariance matrix. Left: constraints from $P_{\rm pre}$, $P_{\rm post}$, $P_{\rm cross}$, and their joint analysis. Right: comparison of $P_{\rm pre}$, $P_{\rm pre} + P_{\rm post}$, and $P_{\rm pre}+P_{\rm post}+P_{\rm cross}$, highlighting the additional information gained by successively including $P_{\rm post}$ and $P_{\rm cross}$. We use $k_{\rm max}=0.18 \,h\,{\rm Mpc}^{-1}$ for $P_{\rm pre}$ and $P_{\rm post}$, and $k_{\rm max}=0.12\,h\,{\rm Mpc}^{-1}$ for $P_{\rm cross}$.}
    \label{fig:difftype}
\end{figure*}

Fig.~\ref{fig:difftype} further illustrates the information gain from the joint analysis using the Gaussian covariance. Here, we adopt $k_{\rm max}=0.18 \,h\,{\rm Mpc}^{-1}$ for $P_{\rm pre}$ and $P_{\rm post}$, and $k_{\rm max}=0.12\,h\,{\rm Mpc}^{-1}$ for $P_{\rm cross}$. The left panel compares the constraints obtained from the individual $P_{\rm pre}$, $P_{\rm post}$, and $P_{\rm cross}$ analyses with those from their joint analysis. The joint analysis clearly yields tighter constraints than any of the individual power spectra. The right panel further shows the additional information provided by $P_{\rm post}$ and $P_{\rm cross}$ relative to $P_{\rm pre}$ alone. Adding $P_{\rm post}$ improve the constraint on $\ln(10^{10}A_s)$ by $33.7\%$, reducing the marginalized error from $0.126$ to $0.084$. Further including $P_{\rm cross}$ improves it by an additional $6.4\%$, yielding an uncertainty of $0.078$. The overall improvement for the $\ln(10^{10}A_s)$ parameter relative to $P_{\rm pre}$ alone is $37.9\%$. For $\Omega_{\rm de}$, adding $P_{\rm post}$ improves the constraint by $21\%$, reducing its marginalized uncertainty from $0.0118$ to $0.0093$. Further including $P_{\rm cross}$ provides an additional $11.3\%$ improvement, yielding an uncertainty of $0.0082$. The overall improvement for the $\Omega_{\rm de}$ parameter relative to $P_{\rm pre}$ alone is $30\%$. The $P_{\rm pre}$-only analysis provides only a weak constraint on $\omega_c$, with the posterior not well bounded, and we therefore do not quote a relative improvement for this parameter. Overall, these results show that most of the additional constraining power comes from $P_{\rm post}$. The inclusion of $P_{\rm cross}$ further improves the joint constraints, although the improvement is modest.

\section{Conclusions} \label{sec:conclusion}

In this work, we apply the Gaussian covariance formalism to develop a semi-analytical Gaussian covariance for the joint analysis of pre-reconstruction, post-reconstruction, and cross full-shape galaxy power spectra. The framework explicitly accounts for the correlations among the pre-and post-reconstructed density fields, and provides a fast alternative to covariance estimation from large suites of mock catalogues. In particular, we model the scale-dependent cross shot noise using the displacement field statistics. We further introduce a new estimator that directly measures the reconstruction-reduced cross shot noise without splitting the galaxy catalogue, thereby slightly reducing the numerical statistical uncertainty of $P_{\rm cross}$ on small scales.

We validate the semi-analytical Gaussian covariance using GLAM mock catalogues. The analytical cross shot-noise model reproduces the scale dependence measured from the mocks, demonstrating that the Gaussian displacement approximation provides an accurate description of the reconstruction-induced noise contribution. 

The semi-analytical Gaussian covariance model reproduces the dominant covariance structure of the individual covariance blocks, including the auto-correlations of each observable and the cross-correlations among different power-spectrum multipoles. For the joint analysis of $P_{\rm pre} + P_{\rm post} + P_{\rm cross}$ with $k_{\rm max}=0.18\,h\,{\rm Mpc}^{-1}$ for $P_{\rm pre}$ and $P_{\rm post}$, and $k_{\rm max}=0.12\,h\,{\rm Mpc}^{-1}$ for $P_{\rm cross}$, the cosmological parameter constraints obtained using the semi-analytical Gaussian covariance matrix are in good agreement with those derived from the HS-HD-based numerical covariance estimated from the GLAM mock catalogues. We further quantify the information gain from the joint analysis using the Gaussian covariance matrix, finding that most of the additional constraining power relative to $P_{\rm pre}$ comes from $P_{\rm post}$, while including $P_{\rm cross}$ provides a modest further improvement over the $P_{\rm pre}+P_{\rm post}$ analysis. Although the parameter inference is performed using the HS-HD method, the variances of $P_{\rm cross}$ obtained with the HS-HD method and the new estimator differ by less than 2\% up to $k \leq 0.12 \,h\,{\rm Mpc}^{-1}$. We therefore expect the choice between the two estimators to have a negligible impact on the cosmological constraints reported here.

The framework explored here enables efficient computation of the covariance matrix for the joint analysis of pre- and post-reconstruction fields in future galaxy surveys. While the Gaussian covariance provides an adequate description over the scales considered in this work, non-Gaussian contributions become increasingly important on smaller scales, particularly for the cross power-spectrum covariance due to the nonlinear coupling between the pre- and post-reconstruction fields. Future improvements will focus on incorporating these non-Gaussian effects, including the trispectrum contribution \cite{Scoccimarro:1999kp,Sugiyama:2019ike} and the super-sample covariance \cite{Hu:2002we,Takada:2013wfa}. Furthermore, for applications to real survey data with irregular geometries, the Gaussian covariance framework can be extended by incorporating survey window effects to account for the survey footprint \cite{Li:2018scc}. These extensions will improve the accuracy of analytical covariance modelling and broaden its applicability to future precision full-shape analyses.

\section*{Acknowledgments}

YW is supported by NSFC Grants (12273048, 12422301), National Key R\&D Program of China No. (2023YFA1607803), and the CAS Project for Young Scientists in Basic Research (No. YSBR-092).  KK is supported also by STFC grant number ST/B001175/1. Numerical computations were performed on the UK Sciama High Performance Computing cluster, supported by the ICG at the University of Portsmouth, and the Tsinghua Astrophysics High-Performance Computing platform at Tsinghua University. For the purpose of open access, we have applied a Creative Commons Attribution (CC BY) licence to any Author Accepted Manuscript version arising. Supporting research data are available on reasonable request.

\bibliography{ms}
\bibliographystyle{jhep}

\end{document}